\RequirePackage{fix-cm}

\documentclass[smallextended]{svjour3}

\smartqed  

\usepackage{graphicx}
\usepackage{mathptmx}      
\usepackage{amssymb}
\usepackage{amsmath}
\usepackage{amsfonts}
\usepackage[round]{natbib}
\usepackage[utf8]{inputenc}
\usepackage[official]{eurosym}
\usepackage[labelfont=bf]{caption}
\usepackage{subcaption} 
\usepackage{float}
\usepackage{authblk}
\usepackage{hyperref}
\usepackage{xcolor}
 
\newbox{\myorcidaffilbox}
\sbox{\myorcidaffilbox}{\large\includegraphics[height=1.7ex]{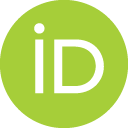}}

\newcommand{\orcidaffil}[1]{%
  \href{https://orcid.org/#1}{\usebox{\myorcidaffilbox}}}

\journalname{Experimental Economics}

\begin{document}

\title{Improving the Power of Economic Experiments Using Adaptive Designs
}


\author{Sebastian Jobjörnsson \and Henning Schaak \and Oliver Mußhoff \and Tim Friede}


\institute{
  Sebastian Jobjörnsson (corresponding author) \at
  Department of Medical Statistics, University Medical Center Göttingen, Göttingen, Germany \orcidaffil{0000-0002-4068-4342} \\
  \email{sebastian.jobjoernsson@med.uni-goettingen.de} \and
  Henning Schaak \at Department of Economics, Swedish University of Agricultural Sciences, Uppsala, Sweden \orcidaffil{0000-0002-7659-4795} \and
  Oliver Mußhoff \at Department of Agricultural Economics and Rural Development, Georg-August-Universität Göttingen, Göttingen, Germany \orcidaffil{0000-0002-3746-623X} \and
  Tim Friede \at Department of Medical Statistics, University Medical Center Göttingen, Göttingen, Germany \orcidaffil{0000-0001-5347-7441}
}

\date{Received: TBD / Accepted: TBD}

\maketitle
  
\begin{abstract}
  An important issue for many economic experiments is how the experimenter can ensure sufficient power for rejecting one or more hypotheses. Here, we apply methods developed mainly within the area of clinical trials for testing multiple hypotheses simultaneously in adaptive, two-stage designs. Our main goal is to illustrate how this approach can be used to improve the power of economic experiments. Having briefly introduced the relevant theory, we perform a simulation study supported by the open source R package \texttt{asd} in order to evaluate the power of some different designs. The simulations show that the power to reject at least one hypothesis can be improved while still ensuring strong control of the overall Type I error probability, and without increasing the total sample size and thus the costs of the study. The derived designs are further illustrated by applying them to two different real-world data sets from experimental economics.
  
  \vspace{0.25cm}
  \hspace{-0.6cm} \textbf{JEL classification:} C12\footnote{C12 = Hypothesis Testing: General.} $\cdot$ C90\footnote{C90 = Design of Experiments: General.} \\
\end{abstract}

\thispagestyle{empty}

\clearpage

\section*{Declarations}

\noindent \textbf{Funding} \\ 
None. \\

\noindent \textbf{Conflicts of interest/Competing interests} \\
None. \\

\noindent \textbf{Availability of data and material} \\
The data used was from the publications by \citet{MusshoffHirschauer2014} and \citet{KarlanList2007}. The data from \citet{MusshoffHirschauer2014} will be made available as supplementary material after review. Data from \citet{KarlanList2007} may be found at \url{https://dataverse.harvard.edu/dataset.xhtml?persistentId=doi\%3A10.7910/DVN/27853}. \\

\noindent \textbf{Code availability} \\
The code will be available at the repository of Experimental Economics, if the article is accepted for publication. \\

\thispagestyle{empty}

\clearpage

\setcounter{page}{1}

\section{Introduction} \label{sec:introduction}
A key question when planning an experiment is how to use the available resources to maximise the probability of demonstrating the presence of an expected effect. A common formal approach is to first express the research questions as one or more null hypotheses, and then choose an experimental design that gives a high statistical power to reject at least a subset of these hypotheses, given certain effect sizes. Ensuring sufficient statistical power has been acknowledged as an important issue in economic research for some time \citep{de_long_are_1992}, and remains a challenge in empirical work \citep{ziliak_size_2004, ioannidis_power_2017}.

In the aftermath of the replication crisis in psychology \citep[see e.g.][]{pashler_editors_2012, open_science_collaboration_estimating_2015} the issue of insufficient power has gained additional interest, particularly in experimental economics where researchers have been encouraged to account for it \citep{canavari_how_2019,list_so_2011, czibor_dozen_2019}. In principle, higher levels of power can be achieved by simply increasing the sample sizes. However, these are often limited in practice. Researchers may face budget or time related constraints, or may have only a small available pool of potential experimental subjects, either due to a small target population or due to restricted access to the wider population. In experimental economics, methods to investigate the required sample sizes to detect hypothesised treatment effects predominantly assume an experiment which is of a fixed size in the planning phase \citep[cf.][]{bellemare_simulating_2016}. 

One way to stay within a pre-defined budget is to keep the total sample size fixed by increasing the number of hypotheses tested in the experiment, while decreasing the number of participants assigned to each. It is also possible to modify the design after one or several interim analyses, in order to focus the remaining resources on investigating the more promising hypotheses. While such adaptive designs have been applied in other fields (such as clinical trials) for decades \citep[cf.][]{Bauer2016}, the concept has only recently become a topic of interest in the economic literature \citep{kasy_adaptive_2019}. When investigating multiple hypotheses (potentially multiple times, in case of an adaptive design), an additional threat to the validity of the final conclusions stems from an inflation of the Type I error probability. This fundamental issue, also referred to as family-wise error rate (FWER) inflation, often arises in experimental research, where multiple treatments are frequently studied \citep{list_multiple_2019}. Still, it has gained the interest of experimental economists only recently \citep{list_multiple_2019, thompson_simple_2019}.

In this contribution, we demonstrate how to use and evaluate adaptive designs in situations where multiple hypotheses are tested, while still keeping the Type I error probability under control. Although our approach is not new from a methodological viewpoint, its application has mostly been in the area of medical statistics, in particular clinical trials. We wish to widen the scope of application by showing how it can be used to design and analyse economic experiments. As previously mentioned, the economic literature covering adaptive designs is currently limited. For a review of applications in other fields, see e.g.\ \citet{Bauer2016}. \citet{bhat_near-optimal_2020} develop an approach to classical randomisation in experiments with a single treatment for situations where the treatment effect is marred by the effects of covariates and where individual subjects arrive sequentially. Similarly, for experiments where subjects in the sample are observed for multiple periods, Xiong et.\ al.\ (2019, working paper ``Optimal Experimental Design for Staggered Rollouts'') develop an algorithm for staggered rollouts, which optimises the share of treated subjects at different time points. With respect to experiments with multiple treatments and a binary outcome, \citet{kasy_adaptive_2019} develop an algorithm for experiments with many data collection rounds. These methods represent rather specialised approaches, tailored to specific settings. In contrast, the framework to be used here is more general and can be applied under a broader set of conditions. It allows for multiple interim analyses, which can be combined in different ways, and is flexible with respect to the applied adaptation rules. Although it allows for an arbitrary number of interim analyses, we will restrict the illustrations to two-stage designs, containing only a single interim analysis. This will simplify our presentation, but is also a reasonable choice in many practical settings. We use available software (the R package \texttt{asd}) to illustrate the principal concepts and show how they can be applied to achieve a more efficient use of resources available in different settings, without having to develop tailor-made algorithms.

To illustrate the methodology, we make use of real data from two experimental studies. The first \citep{MusshoffHirschauer2014} uses a framed field experiment, based on a business simulation game, to study the effectiveness of different nitrogen extensification schemes in agricultural production. The experiment was carried out with 190 university students and contained treatments which varied in two dimensions: whether nitrogen limits were voluntary or mandatory, and whether the corresponding payments were deterministic or stochastic. The second example \citep{KarlanList2007} uses a natural field experiment to study the effects of different matching grants on charitable donations. The experiment targeted previous donors of a non-profit organisation, and considered treatments which varied in three different dimensions (the matching ratio, the maximum size of the matching grant and the suggested donation amount). The original study is based on a large sample, including 50,083 individuals. In order to simplify the presentation and keep the focus on our core topics, we will only include a subset of the data in our analyses. More detailed descriptions of the experiments corresponding to our two applications may be found in Appendix \ref{sec:appendix1} and \ref{sec:appendix2}. 

The remaining content of this paper is structured as follows. Section \ref{sec:theory} introduces the basic theory and notation for the multiple testing procedures and adaptive designs considered. Section \ref{sec:simulation} then proceeds with demonstrating how to evaluate the power properties of adaptive designs with a single interim analysis via simulation using the freely available R package \texttt{asd}. A more detailed description of this simulation study, providing experimental instructions for those who which to reproduce our results, is given in Appendix \ref{sec:appendix3}. In Section \ref{sec:applications}, we apply specific two-stage designs to the two different data sets mentioned above and compare their performance to more traditional single-stage designs with the same total sample size. Section \ref{sec:discussion} concludes with a discussion of more advanced variations of adaptive designs further explored in other parts of the literature.

\section{Type I error probability control for multiple hypotheses in adaptive designs} \label{sec:theory}

\subsection{Multiple testing procedures} \label{sec:multiple_testing_procedures}
As mentioned in the introduction, one way to increase the power to find \emph{some} result of interest is simply to test more hypotheses in the same study. However, unless a proper multiple testing procedure is used, increasing the number of hypotheses tested typically leads to a larger Type I error probability. This is problematic since a goal of many designs is to keep this error under strict control at a certain pre-specified level. The phenomenon can be illustrated by the following simple example. Suppose that $m$ different null hypotheses are tested based on independent test statistics, and assume that the tests used have been chosen so as to make the individual (i.e., the nominal) Type I error probabilities equal to $\alpha$. This means that the marginal probability to \emph{not} reject a specific null hypotheses equals $1 - \alpha$, given that it is true and there is no non-zero effect. By the independence assumption, if all null hypotheses are true, then it follows that the probability to reject at least one of them is $1 - (1 - \alpha)^m$. Hence, as $m$ increases, the probability to falsely reject at least one null hypotheses converges to certainty.

To what extent it is important to keep the Type I error probability under strict control, and, if so, which significance level to choose, varies across scientific disciplines and applications. In order to choose appropriate values of the Type I error probability and power, one must weigh the value of correctly finding a positive effect against the loss of falsely declaring an effect as positive when it is not. For example, in the regulated area of confirmatory clinical trials for licensing new medical treatments, a Type I error probability of $0.025$ or $0.05$ is typically required by the authorities (for one-sided and two-sided hypotheses, respectively). In fact, since typically two independent studies would be required to successfully demonstrate an effect, this means that the Type I error probability would actually be much smaller across the studies. If the effect is assumed to be small, this can lead to large and costly studies. Nevertheless, it is often easy to motivate the importance of keeping the Type I error probability this small, since incorrectly allowing non-working medicines on the market could potentially lead to severe negative health effects. Here, our interest does not lie in specifying certain Type I error probabilities as appropriate, or even to argue that they should always be controlled. We assume that this has been deemed appropriate, and aim to demonstrate the methods that, to a large extent, have been developed within the field of medical statistics.

Let $H_1, \ldots, H_m$ denote a set of $m$ null hypotheses of interest, $m_0$ of which are true. Before proceeding to discuss multiple testing procedures, note that there have been several different suggestions of how to generalise the concept of a Type I error probability for a single hypothesis to the case of multiple hypotheses. These are referred to as different \emph{error rates} (see, e.g., Chapter 2 of \citet{BretzHothornWestfall2011}). For example, the \emph{per comparison error rate} is the expected proportion of falsely rejected hypotheses and equals $m_0 \alpha / m$ if each hypothesis is tested individually at level $\alpha$. Here, we will only focus on controlling the \emph{family-wise error rate} (FWER), defined as the probability of making at least one Type I error. This probability depends on the specific combination of hypotheses that are actually true, which of course is unknown when planning the experiment.

In order to give a detailed account of multiple testing procedures for controlling the FWER, we first need to consider the concept of intersection hypotheses. Given $m$ individual hypotheses $H_1, \ldots, H_m$ and a non-empty subset of indices $I \subseteq \{ 1, \ldots, m \}$, the intersection hypothesis $H_I$ is defined as
\begin{equation} \label{eq:intersection_hypothesis}
  H_I = \bigcap_{i \in I} H_i, \quad I \subseteq \{1, \ldots, m\}. 
\end{equation}
\emph{Local} control of the FWER at level $\alpha$ for a specific intersection hypothesis $H_I$ holds when the conditional probability to reject at least one hypothesis given $H_I$ is at most $\alpha$, i.e.\ when
\begin{equation} \label{eq:local_FWER_control}
  \mathbb{P}_{H_I} \left(  \text{Reject $H_i$ for some $i \in I$} \right) \le \alpha. 
\end{equation}
\emph{Strong} control of the FWER means that the inequality in Eq.\ \eqref{eq:local_FWER_control} must hold for \emph{all} possible non-empty subsets $I$. This more conservative requirement bounds the probability of making at least one Type I error regardless of which hypotheses are true. It is often deemed appropriate when the number of hypotheses is relatively small, while the consequences of making an error may be severe.

Now, given that tests have been defined for the local control of a collection of intersection hypotheses, how can this be used to attain strong control of the FWER? One option is to employ the widely used \emph{closed testing principle}. This principle states that if we reject an individual hypothesis $H_i$ if and only if all intersection hypotheses $H_I$ such that $i \in I$ are rejected at local level $\alpha$, then the FWER is strongly controlled at level $\alpha$ \citep{Marcus1976}. Since this principle is completely general as regards the specific form of the tests for the intersection hypotheses $H_I$, it follows that the choice of local tests will determine the properties of the overall procedure. 

There are a number of different multiple testing procedures available for local control of the FWER. One of the most widely applicable is the well-known Bonferroni procedure, due to the fact that no distributional assumptions are required for the statistical model. In this procedure, it is first assumed that $m$ individual nominal tests have been defined for the hypotheses $H_1, \ldots, H_m$, say based on ordinary t-tests. The Bonferroni procedure then rejects $H_i$ at level $\alpha$ if the corresponding nominal test would have rejected $H_i$ at level $\alpha / m$. That this procedure locally controls the FWER for an intersection hypothesis $H_I$ follows immediately from the Bonferroni inequality, since
\begin{equation} \label{eq:local_Bonferroni}
  \mathbb{P}_{H_I} \left(  \text{Reject $H_i$ for some $i \in I$} \right) \le \sum_{i \in I} \mathbb{P}_{H_I} \left(  \text{Reject $H_i$} \right) \le \frac{\left| I \right| \alpha}{ m } \le \alpha . 
\end{equation}

In the following, we will use the Dunnett procedure \citep{Dunnett1955} for testing intersection hypotheses. This test is tailored to our specific situation, in which several different active treatments are compared to a common control. Specifically, it is assumed that each individual observation belongs to one of $m + 1$ different treatment groups. The group with index $i = 0$ is the control group, against which the other groups are compared. The individual observations in each group are used to form group means $\bar{X}_i$. Due to the central limit theorem, we assume these to be normally distributed, with true means $\mu_i$, $i = 0, \ldots, m$ and variances $\sigma^2 / n$, where $\sigma^2$ is assumed known and $n$ is the group size (all assumed equal). We thus have  
\begin{equation} \label{eq:sample_means}
  \bar{X}_{i} \sim \textrm{N} \left( \mu_i, \sigma^2 / n \right), \quad i = 0, 1, \ldots, m . 
\end{equation}

After the experiment, these observations yield the estimates $\bar{X}_i - \bar{X}_0$ of the treatment mean differences relative to control. The actual testing is based on the standardised mean differences,
\begin{equation} \label{eq:Z_values}
  Z_i = \frac{\bar{X}_i - \bar{X}_0}{\sqrt{2 \sigma^2 / n}} \sim \textrm{N} \left( \frac{\mu_i - \mu_0}{\sqrt{2 \sigma^2 / n}}, 1 \right), \quad i = 1, \ldots, m.
\end{equation}
These assumptions lead to a certain covariance structure for the joint distribution of the statistics $Z_1, \ldots, Z_m$ which is exploited by the Dunnett test. The individual null hypotheses are $H_i : \mu_i - \mu_0 = 0$. For a given intersection hypothesis $H_I$, the test statistic is defined as
\begin{equation} \label{eq:Z_max_test_stat}
  Z_{I}^{\max} = \max_{i \in I} Z_i .
\end{equation}
Given that a specific value $Z_{I}^{\max} = z$ has been observed, a corresponding p-value may be computed as $p_{I} = 1 - F_{I}^{\max} (z)$ \citep{FriedeStallard2008}, where
\begin{equation} \label{eq:Z_max_dist}
F_{I}^{\max} (z) = 1 - \int_{- \infty}^\infty \left[ \Phi \left( \sqrt{2} z + x \right) \right]^{|I|} \phi (x) \, \mathrm{d} x
\end{equation}
is the cumulative distribution function of $Z_{I}^{\max}$, $\phi$ the standard normal density and $\Phi$ the standard normal distribution function. The intersection hypothesis $H_I$ is rejected at level $\alpha$ if $p_{I} \le \alpha$. Due to the closed testing principle, the individual hypothesis $H_i$ may be rejected while strongly controlling the FWER if $p_I \le \alpha$ for all index sets $I$ such that $i \in I$.

\subsection{Adaptive designs} \label{sec:adaptive_designs}
If we can look at some of the data before spending all available resources, then we could allocate the remaining samples to the most promising treatments. The present section contains a short introduction to the general topic of such adaptive designs, sufficient for our forthcoming applications. In particular, we focus here on the developments made by researchers in the area of medical statistics and its application to clinical trials, which has driven much of the methodological progress since the early 1990s. As mentioned in the introduction, for reasons of clarity, we will limit ourselves to the case of two-stage designs with a single interim analysis. There are two main approaches. One is referred to as the \emph{combination test} approach and was introduced by \citet{Bauer1989} and \citet{BauerKohne1994}. The other, which makes use of \emph{conditional error functions}, was introduced by \citet{ProschanHunsberger1995} for sample size recalculation, while \citet{MuellerSchaefer2004} widened the approach to allow for general design changes during any part of the trial.

Following \citet{WassmerBrannath2016}, assume for the moment that the purpose of the study is to test a single null hypothesis $H_0$. We will see later how to combine interim adaptations with the closed testing principle when several hypotheses are involved. When the interim analysis is performed, there are two possibilities. Either $H_0$ is rejected solely based on the interim data $T_1$ from the first stage or the study continues to the second stage, yielding data that we assume can be summarised by means of a statistic $T_2$. In general, the null distribution of $T_2$ may then depend on the interim data. For instance, the first stage could inform an adjustment of the sample size of the second stage. Nevertheless, it is often the case that $T_2$ may be transformed so as to make the resulting ditribution invariant with respect to the interim data and the adjustment procedure. In other words, the statistics used to evaluate the two stages will be independent given that $H_0$ is true, regardless of the interim data and the adaptive procedure.

As an example, suppose that $T_2$ is a normally distributed sample mean under $H_0$, with mean $0$ and a known variance $\sigma^2 / n_2$, with $\sigma^2$ being the variance of a single observation and $n_2$ a second stage sample size that depends on the interim data in some way. Then a one-sided p-value for the second stage can be obtained by standardising $T_2$ according to
\begin{equation}
  p_2 = 1 - \Phi \left( \frac{T_2}{ \sigma / \sqrt{n_2} } \right).
\end{equation}
Assuming that the data from the two different stages is independent, it follows that the joint distribution of the p-values $p_1$ and $p_2$ from the two stages is known and invariant with respect to the interim adjustment. This is the key property which allows for the flexibility when choosing the adaptive design procedure.

Based on the conditional invariance principle, the design of an adaptive trial controlling the Type I error probability at a level $\alpha$ can be specified in terms of a rejection region for $p_1$ and a corresponding region for $p_2$. This results in a sequential test of $H_0$ which is independent of the adaptation rule. The combination test approach and conditional error function approach only differ in the way the rejection region for the second stage is specified. Suppose that a one-sided null hypothesis $H_0 : \theta \le 0$, where $\theta$ is some parameter of interest, is to be tested using a two-stage design with p-values $p_1$ and $p_2$. The combination test procedure is specified in terms of three real-valued parameters, $\alpha_0$, $\alpha_1$ and $c$, and a \emph{combination function} $C(p_1, p_2)$. It proceeds as follows:
\begin{enumerate}
\item After the interim analysis, stop the trial with rejection of $H_0$ if $p_1 \le \alpha_1$ and with acceptance of $H_0$ if $p_1 > \alpha_0$. Otherwise, apply the adaptation rule and collect the data of the second stage.
\item After the second stage, reject $H_0$ if $C(p_1, p_2) \le c$ and accept $H_0$ if $C(p_1, p_2) > c$.
\end{enumerate}
By virtue of the conditional invariance principle, it is easily shown that the Type I error probability is controlled at level $\alpha$ by any choice of $\alpha_0$, $\alpha_1$, $c$, and $C(p_1, p_2)$ satisfying
\begin{equation} \label{eq:comb_fun_criterion}
   \mathbb{P}_{H_0} \left(  \text{Reject $H_0$} \right) = \alpha_1 + \int_{\alpha_1}^{\alpha_0} \! \int_{0}^{1} \! \mathbf{1} \left\{ C(p_1, p_2) \le c \right\} \, \mathrm{d} p_2 \, \mathrm{d} p_1 \le \alpha ,
\end{equation}
where $\mathbf{1} \left\{ \cdot\right\}$ denotes the indicator function.

The conditional error function approach is very similar. The parameters $\alpha_0$ and $\alpha_1$ have the same role, but $c$ and $C(p_1, p_2)$ are replaced by a \emph{conditional error function} $A(p_1)$ such that, after the second stage, $H_0$ is rejected if $p_2 \le A(p_1)$ and accepted if $p_2 > A(p_1)$. It is important to note that, as long as one of the procedures described above is followed, \emph{any} type of rule can be used to update the sample size in an interim analysis. This means that not only can the information on efficacy collected so far be used in an arbitrary manner, but so also can any information external to the trial. For a more comprehensive review of the history and current status of adaptive designs as applied to clinical trials, see \citet{Bauer2016}.

Although many alternative choices exist, in the designs considered in the following sections we will exclusively make use of the \emph{inverse normal combination function} \citep{LehmacherWassmer1999} when combining the p-values from the two stages. The definition of this function is
\begin{equation} \label{eq:inverse_normal_comb_fun}
  C(p_1, p_2) = 1 - \Phi \left( w_1 \Phi^{-1} (1 - p_1) + w_2 \Phi^{-1} (1 - p_2) \right),
\end{equation}
where $w_1$ and $w_2$ are pre-specified weights that satisfy the requirement $w_1^2 + w_2^2 = 1$. In our applications, we will use the weights suggested by \citet{Jenkins2011}, i.e., $w_1 = \sqrt{n_1 / (n_1 + n_2)}$ and $w_2 = \sqrt{n_2 / (n_1 + n_2)}$, where $n_1$ and $n_2$ are the group sizes for stage 1 and 2, respectively.\footnote{Note that $n_1$ for a two-stage designs corresponds to $n$ for a single-stage design, as used in Eq.\ \eqref{eq:sample_means}.} With $p_1$ and $p_2$ independent and uniformly distributed, it follows from the definition that $C(p_1, p_2)$ becomes a random variable of the form $C(p_1, p_2) = 1 - \Phi(Z)$, where $Z$ follows a standard normal distribution. Thus, $C(p_1, p_2)$ can be treated as a p-value that combines the information from the two stages. In particular, if the first stage parameters of Eq.\ \eqref{eq:comb_fun_criterion} are set to $\alpha_1 = 0$ and $\alpha_0 = 1$, then there will be no early stopping and the null hypothesis will be rejected after the second stage if and only if $C(p_1, p_2) \le \alpha$.

Having briefly introduced the necessary components, we are now ready to describe the step-by-step procedure characterising the class of adaptive designs that will be investigated using simulation in the following section.
\begin{enumerate}
\item Define a set of one-sided, individual null hypotheses $H_1, \ldots, H_m$ corresponding to comparisons of treatment means $\mu_i$ against a common control $\mu_0$ (i.e., $H_i : \mu_i \le \mu_0$). Assume the model given by Eq.\ \eqref{eq:sample_means}.
\item Given the first stage data, apply Dunnett's test and compute a p-value $p_1^I$ for each non-empty intersection hypothesis $H_I$ such that $I \subseteq \{1, \ldots, m\}$.
\item Apply a treatment selection procedure to the first stage data, for example by selecting a fixed number of the best treatments. This results in a subset $J \subseteq \{1, \ldots, m\}$ taken forward to the second stage.
\item Given the second stage data, apply Dunnett's test and compute a p-value $p_2^I$ for each non-empty intersection hypothesis $H_I$ such that $I \subseteq J$.
\item Compute combined p-values $p_c^I = C(p_1^I, p_2^I)$ using a normal combination function for each $I \subseteq J$. Reject each $H_I$ at local level $\alpha$ for which $p_c^I \le \alpha$. Finally, ensure strong control of the FWER at level $\alpha$ by applying the closed testing principle and rejecting the individual hypotheses $H_i$ satisfying $p_c^I \le \alpha$ for all $I \subseteq J$ such that $i \in I$. 
\end{enumerate}
Note that this procedure assumes that those hypotheses dropped at the interim cannot be rejected in the final analysis. This leads to a conservative procedure with an actual Type I error probability that may be lower than the pre-specified level $\alpha$ \citep{Posch2005}.

\section{Design of adaptive experiments using simulation} \label{sec:simulation}

In this section we will apply the adaptive design theory introduced above and demonstrate how available software can be used to evaluate different designs. The absence of user friendly tools can often be a hurdle in practical applications, and while there are commercially available alternatives (e.g., ADDPLAN and EAST), we will make use of the open source R package \texttt{asd}. This simulation package was originally developed for evaluating two-stage adaptive treatment selection designs using Dunnett's test \citep{Parsons2012}, and was later extended to also support adaptive subgroup selection \citep{Friede2020}. Another R package called \texttt{rpact} implements the methods described in the monograph by \citet{WassmerBrannath2016} and might be of interest to readers looking to apply the methodology, but is not used here. In addition, a preprint that will shortly be available on arXiv describes another software package that might be of interest (Richter et. al., 2020, ``Improving Adaptive Seamless Designs Through Bayesian Optimization'').

The \texttt{asd} package will be used to implement the simulations for the step-by-step procedure described at the end of Section \ref{sec:theory}. Mirroring the experimental applications, we will consider two different cases. The first is that of a small total sample size and a large effect, while the second instead considers the case of a large total sample size and a small effect. The \texttt{asd} package automatically handles the issue of strong control of the FWER. The steps we go through can be seen as a design procedure for selecting a two-stage design based primarily on the power to reject at least one hypothesis, and to a lesser extent on the treatment selection probabilities at the interim analysis. However, we stop short of converting this into a precise numerical optimisation problem and instead use a more qualitative approach based on plots. We find this approach to be more instructive for illustrating how packages like \texttt{asd} can be used in general. A full numerical optimisation problem can be derived and solved once the software supporting our procedure is in place (conceptually, at least, although implementation of efficient software can always be a challenge).\footnote{All plots in this section were created using the R code accompanying this paper.} A more detailed description of how de simulation results were obtained may be found in Appendix \ref{sec:appendix3}.

In order to match the assumption of a fixed total budget, all two-stage designs considered have a fixed total sample size. Further, there is also a fixed number of active treatment groups in the first stage. The number of groups in the second stage depends on the selection rule chosen. For simplicity, we will here only consider selection rules $s$ that takes a fixed number of treatments forward from the first to the second stage, where $s = 1$ means that the single best treatment is taken forward, $s = 2$ that the two best are taken forward etc. The groups in each stage are assumed to be of equal size, which implies that two parameters fully characterise the adaptive designs: the selection rule $s$ used at the interim analysis and the proportion of the total sample size used in the first stage. This proportion will henceforth be referred to as the \emph{ratio} and denoted by $r$. The two parameters will be selected using the following procedure:
\begin{enumerate}
\item Specify a total sample size for the trial, and an \emph{expected effect size} $\delta^*$. 
\item Plot the power to reject at least one hypothesis as a function of effect size, for different combinations of the selection rule $s$ and the ratio $r$. Choose the design which corresponds to the highest power in a neighbourhood of $\delta^*$.
\item For the specified value of $\delta^*$, fix the selection rule and ratio found in the previous step and plot the power and selection probabilities in a neighbourhood of the ratio. Adjust the ratio slightly if it leads to better selection probabilities without sacrificing too much power, say at most 1 or 2 percentage points.
\end{enumerate}

The power is always evaluated relative to a specific configuration of the treatment effects. We will consider two different configurations. In the first, the effects are increasing from $0$ to $\delta$ in steps of equal length. In this configuration, we thus have
\begin{equation} \label{eq:effect_config1}
  \mu_i = i \delta / m, \quad i = 0, \ldots, m.
\end{equation}
In the second configuration, the effect sizes of the active treatments are clustered close to $\delta$. Specifically, it is assumed that
\begin{align} 
  \mu_0 & = 0, \label{eq:effect_config2a} \\
  \mu_i & = \delta - (m - i) \eta, \quad i = 1, \ldots, m, \label{eq:effect_config2b}
\end{align}
for some small fixed number $\eta$.

\subsection{Two-stage design for a small sample and a large effect} \label{sec:design_small_sample_large_effect}
In this section we search for a good two-stage adaptive design with two active treatment groups in the first stage and a small total sample size of $100$. The expected effect size is assumed to be $\delta^* = 0.5$ and our primary aim is to find the design with the highest power in a neighbourhood of this effect. In particular, we would like to see some improvement when comparing against a single-stage design with the same total sample size. Throughout this section and the next, it is assumed that the Type I error probability is controlled at the one-sided level $\alpha = 0.025$.

\begin{figure}
  \centering
  \begin{subfigure}[b]{0.49\textwidth}
    \centering
    \includegraphics[width=\textwidth]{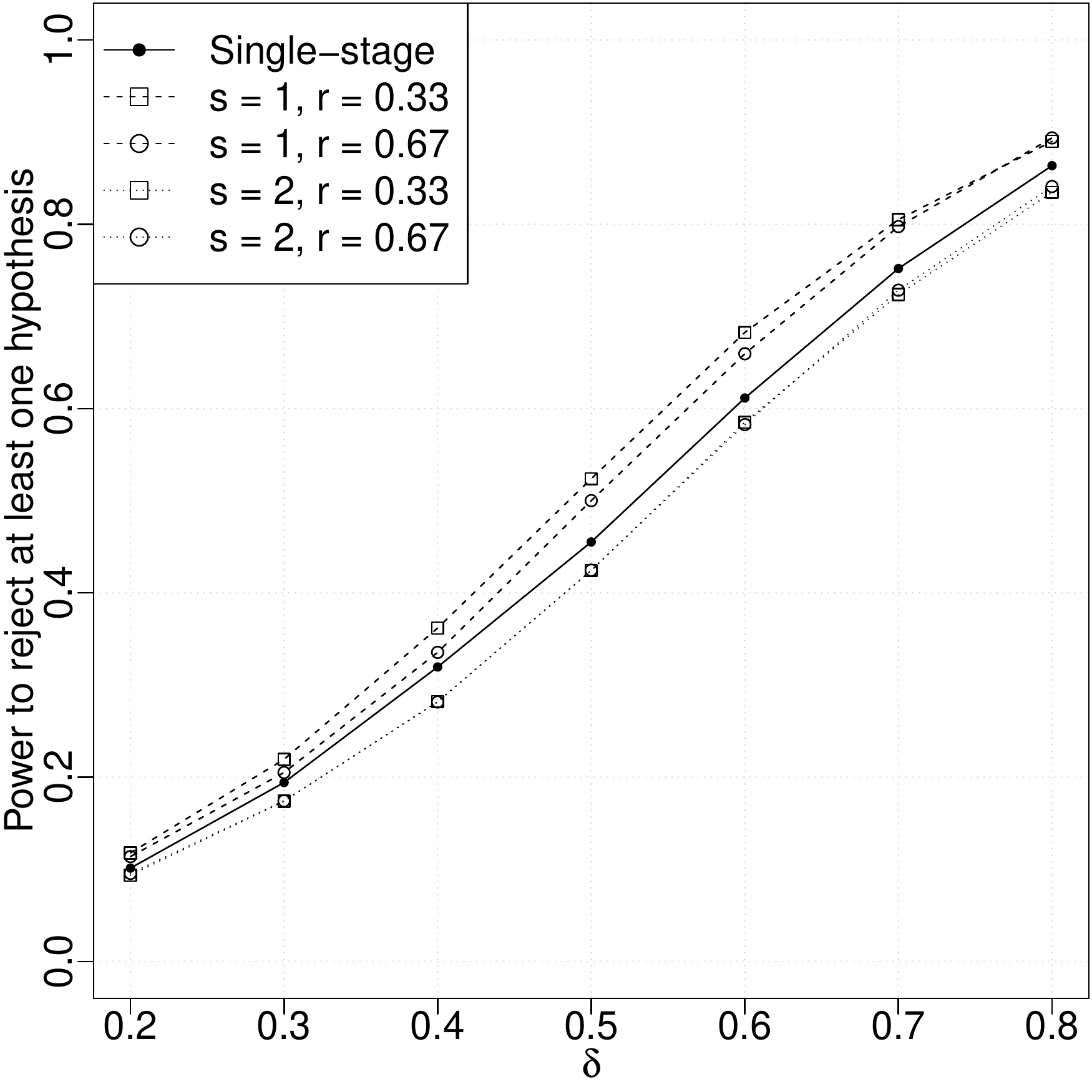} 
    \subcaption{\small{Effect sizes: $0$, $\delta / 2$, $\delta$}}
    \label{fig:power_linear1}
  \end{subfigure}
  \hfill
  \begin{subfigure}[b]{0.49\textwidth}
    \centering
    \includegraphics[width=\textwidth]{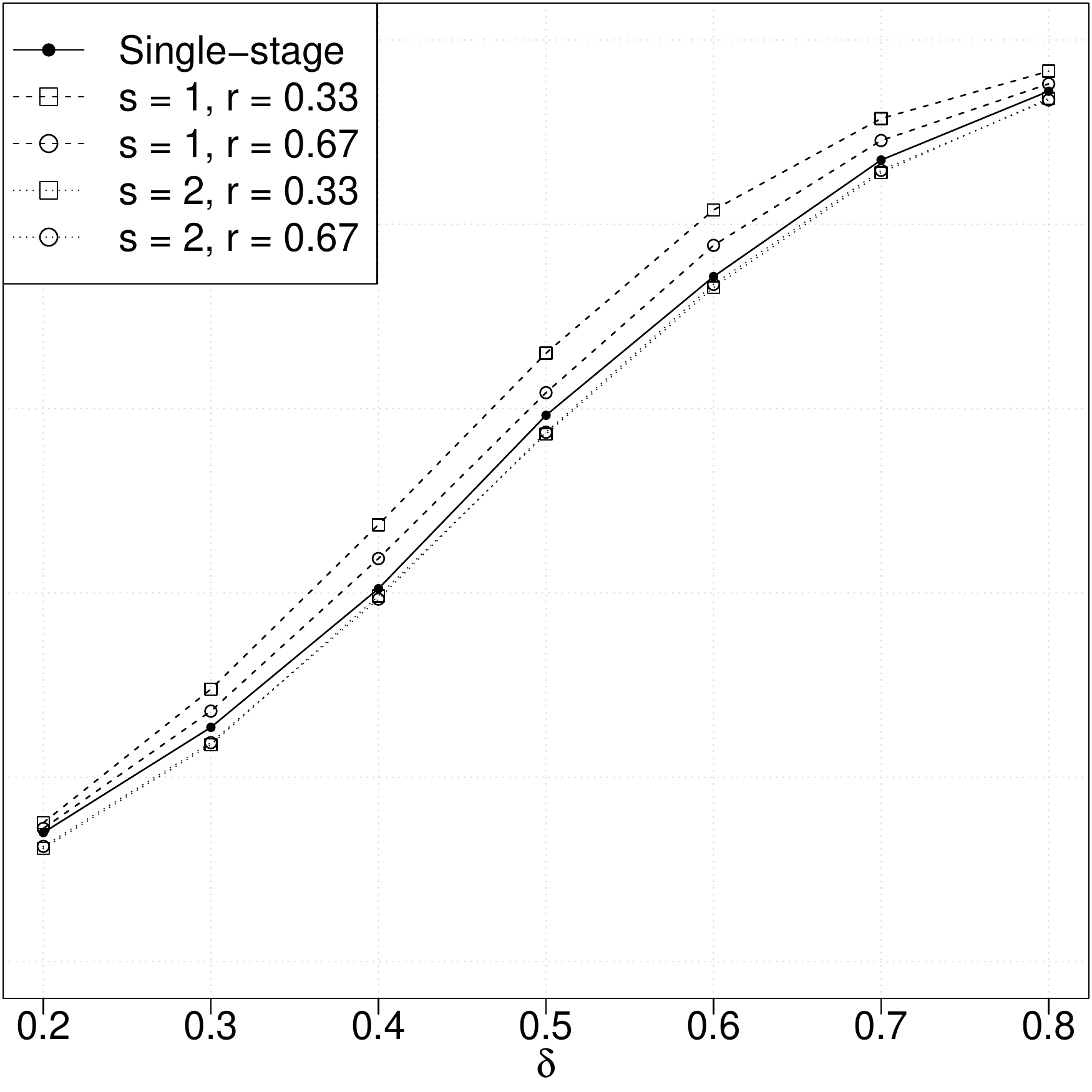} 
    \subcaption{\small{Effect sizes: $0$, $\delta - \eta$, $\delta$, with $\eta = 0.01$}}
    \label{fig:power_similar1}
  \end{subfigure}
  \caption{\small{Power to reject at least one hypothesis as a function of the effect size parameter $\delta$, for two different effect configurations given by Eqs.\ \eqref{eq:effect_config1}, \eqref{eq:effect_config2a} and \eqref{eq:effect_config2b}. Two different selection rules are compared. Either the single best treatment is taken forward to the second stage ($s = 1$), or the two best treatments are taken forward ($s = 2$). There are two options for the proportion of the total sample size used in the first stage ($r = 0.33$ or $r = 0.67$)}}
  \label{fig:power1}
\end{figure}

Figure \ref{fig:power_linear1} shows the probability to reject at least one hypothesis as a function of $\delta$ for the first type of effect configuration. In addition, it also includes the power of a single-stage design of the same total sample size (solid line). It can be seen that the rule $s = 1$ which selects the single best treatment at interim dominates the single-stage design over the entire range $\delta \in [0.2, 0.8]$, for both values of the ratio $r$. The reason for this is that using $s = 1$ allows for increasing the power by dropping the inferior treatment (with effect $\delta / 2$) at the interim, and focus the remaining resources on the superior treatment (with effect $\delta$). Further, the single-stage design in turn dominates the rule $s = 2$ over the entire range $\delta \in [0.2, 0.8]$. This is not surprising, since using the rule $s = 2$ when we actually only have two active treatments will never lead to a treatment being dropped at interim. This means that there can be no power gain from allocating more resources to the most promising treatment. However, when moving from a single-stage to a two-stage design, there will be a loss in power stemming from the adjustments necessary to keep the Type I error probability under control.

For $s = 2$, it can be seen that the power curves for $r = 0.33$ and $r = 0.67$ are almost equal over the entire range of $\delta$. The small difference observed is only due to the finite samples used in the simulations. The reason is that the inverse normal combination functions, as defined by Eq.\ \eqref{eq:inverse_normal_comb_fun}, leads to combined p-values that are invariant in the ratio as long as no hypotheses are dropped at interim. When $s = 1$, it can be observed that for all values of $\delta$ except the very largest (close to $0.8$), a ratio value of 0.33 dominates the alternative of 0.67. Hence, based purely on power considerations, we are led to consider a design which selects the best treatment at interim, having a ratio value close to $0.33$.

Figure \ref{fig:power_similar1} shows the probability to reject at least one hypothesis as a function of $\delta$ for the second type of effect configuration. For this configuration the design corresponding to $s = 1$ and $r = 0.33$ clearly dominates the others. As in Figure \ref{fig:power_linear1}, the single-stage design does not reach the power for the designs corresponding to $s = 1$, while it is uniformly better than the designs corresponding to $s = 2$. When comparing to Figure \ref{fig:power_linear1}, we can also note that the power is uniformly larger for all curves. The reason is of course that a specific value of $\delta$ yields equal effects for the two best treatments, while treatments that are second best have effects $\delta - \eta$ and $\delta / 2$, respectively. Finally, we note that the power curves corresponding to $s = 2$ are much closer to the single-stage design in Figure \ref{fig:power_similar1}. Again, as for the first effect configuration, we are led to consider a design which selects the best treatment at interim, having a ratio value around $0.33$.

\begin{figure}
  \centering
  \begin{subfigure}[b]{0.49\textwidth}
    \centering
    \includegraphics[width=\textwidth]{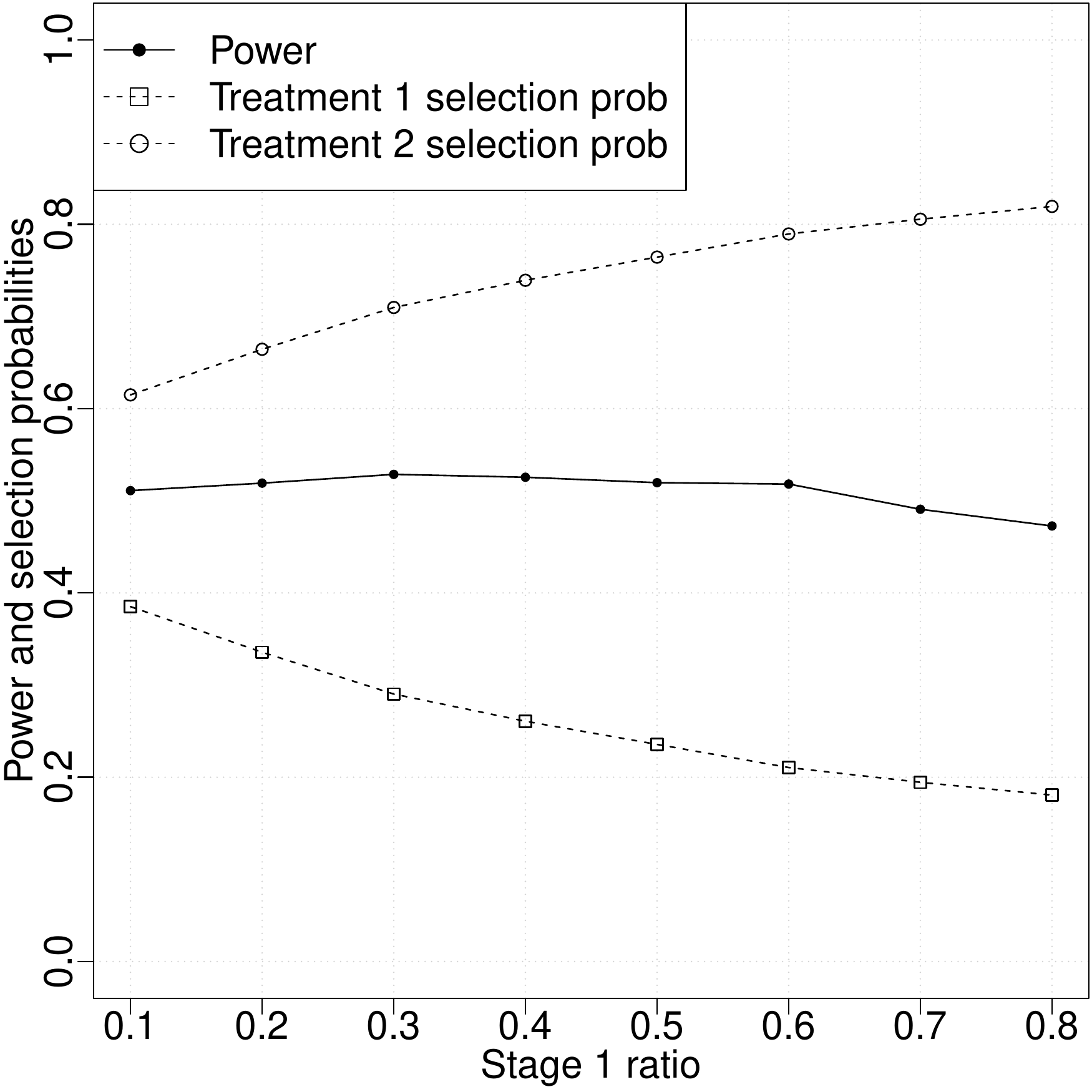} 
    \subcaption{\small{Effect sizes: $0$, $\delta / 2$, $\delta$, with $\delta = 0.5$}}
    \label{fig:selection_probs_linear1}
  \end{subfigure}
  \hfill
  \begin{subfigure}[b]{0.49\textwidth}
    \centering
    \includegraphics[width=\textwidth]{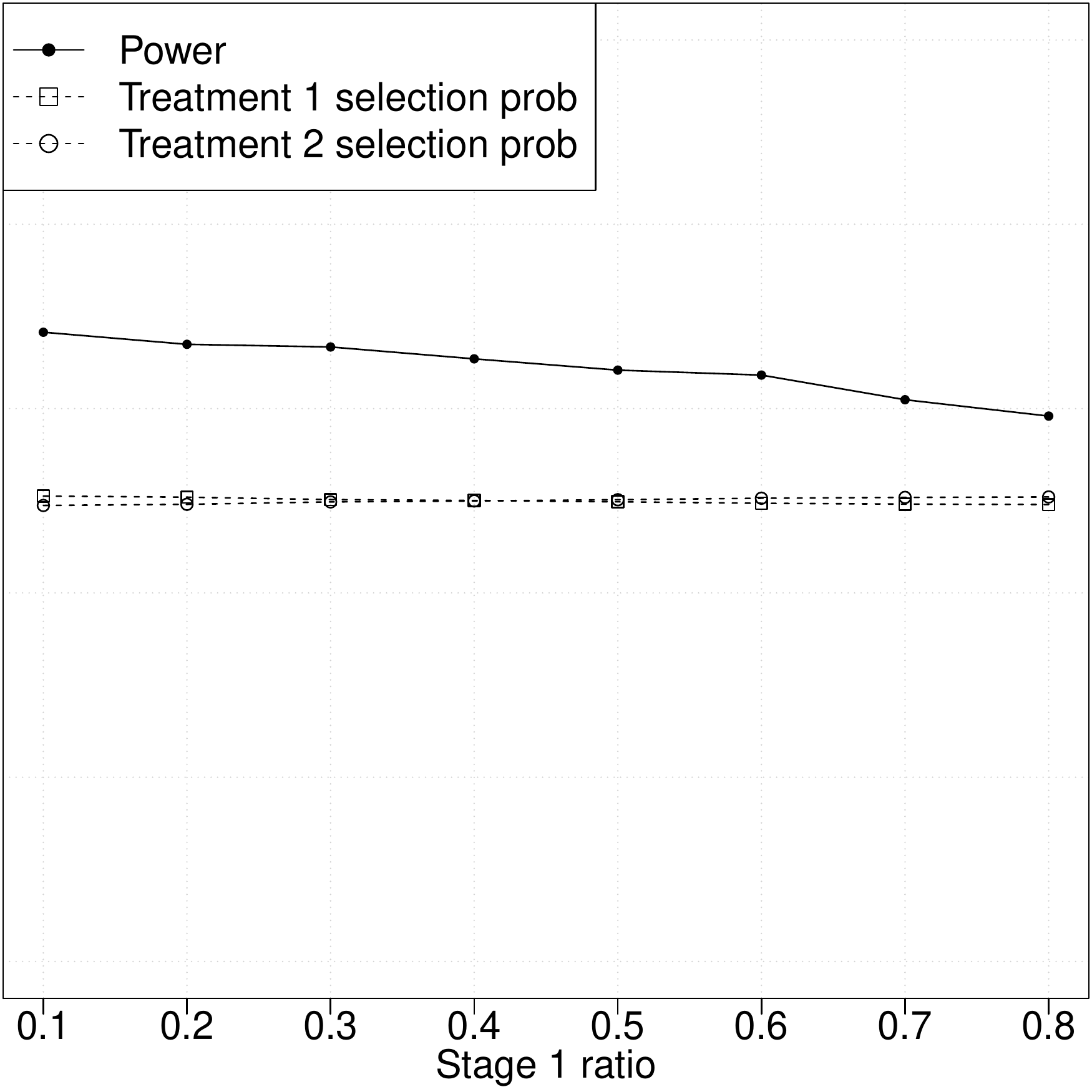} 
    \subcaption{\small{Effect sizes: $0$, $\delta - \eta$, $\delta$, with $\delta = 0.5, \eta = 0.01$}}
    \label{fig:selection_probs_similar1}
  \end{subfigure}
  \caption{\small{Power to reject at least one hypothesis and selection probabilities as functions of the ratio for a two-stage design which selects the single best active treatment at interim, for the two different effect configurations given by Eqs.\ \eqref{eq:effect_config1}, \eqref{eq:effect_config2a} and \eqref{eq:effect_config2b}}}
  \label{fig:selection_probs1}
\end{figure}

Next, consider Figure \ref{fig:selection_probs_linear1}, which shows the power and selection probabilities at the interim analysis as functions of the ratio $r$ for the effect configuration given by Eq.\ \eqref{eq:effect_config1}. The selection rule is now fixed to be $s = 1$ and the value of $\delta$ is fixed at $\delta = \delta^* = 0.5$. It can be seen that the power stays approximately constant at a value just above $0.5$ until $r$ approaches $0.6$. This implies that we can increase the probability of selecting the best treatment at stage 1 without affecting the power too much by increasing the value of the ratio from $0.33$ to, say, $0.6$. 

Finally, consider Figure \ref{fig:selection_probs_similar1}, corresponding to the effect configuration given by Eqs.\ \eqref{eq:effect_config2a} and \eqref{eq:effect_config2b}. As before, the value of $\delta$ is fixed at $0.5$. Since this effect configuration gives almost equal effects (namely, $\delta - \eta$ and $\delta$), it is no suprise that the treatment selection probabilities are almost equal. Of more interest is the fact that the power is monotonically decreasing. The reason is again the similar effect sizes of the two treatments, which implies that the information obtained in the first stage is essentially worthless when selecting which treatment to take forward. Consequently, the resources are better spent on selecting a treatment as soon as possible and then use all resources to improve the estimate in the second stage. We can therefore conclude that a small value of the ratio should be used for this type of effect configuration.

\subsection{Two-stage design for a large sample and a small effect} \label{sec:design_large_sample_small_effect}
\begin{figure}
  \centering
  \begin{subfigure}[b]{0.49\textwidth}
    \centering
    \includegraphics[width=\textwidth]{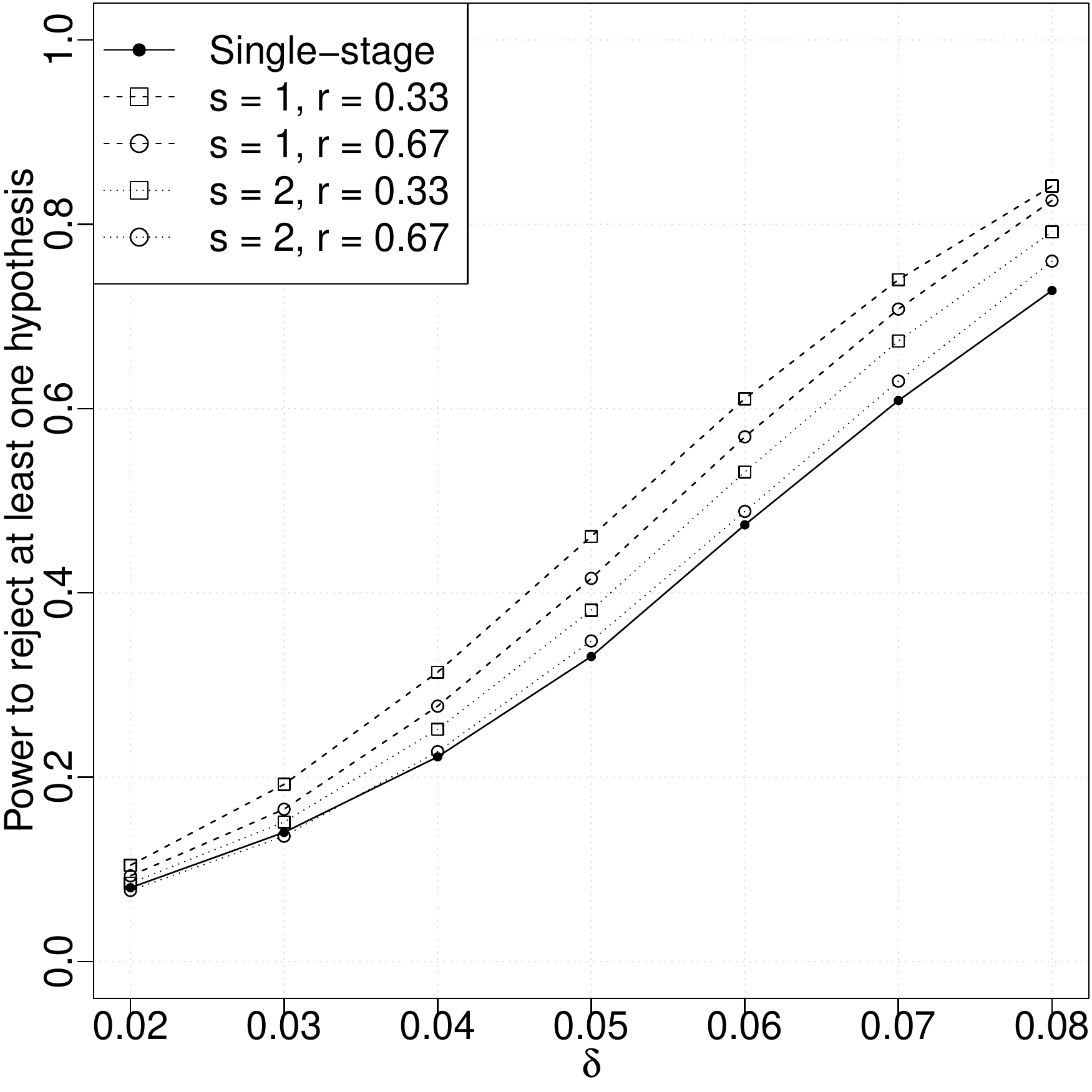} 
    \subcaption{\small{Effect sizes: $0$, $\delta / 3$, $2 \delta / 3$, $\delta$}}
    \label{fig:power_linear2}
  \end{subfigure}
  \hfill
  \begin{subfigure}[b]{0.49\textwidth}
    \centering
      \includegraphics[width=\textwidth]{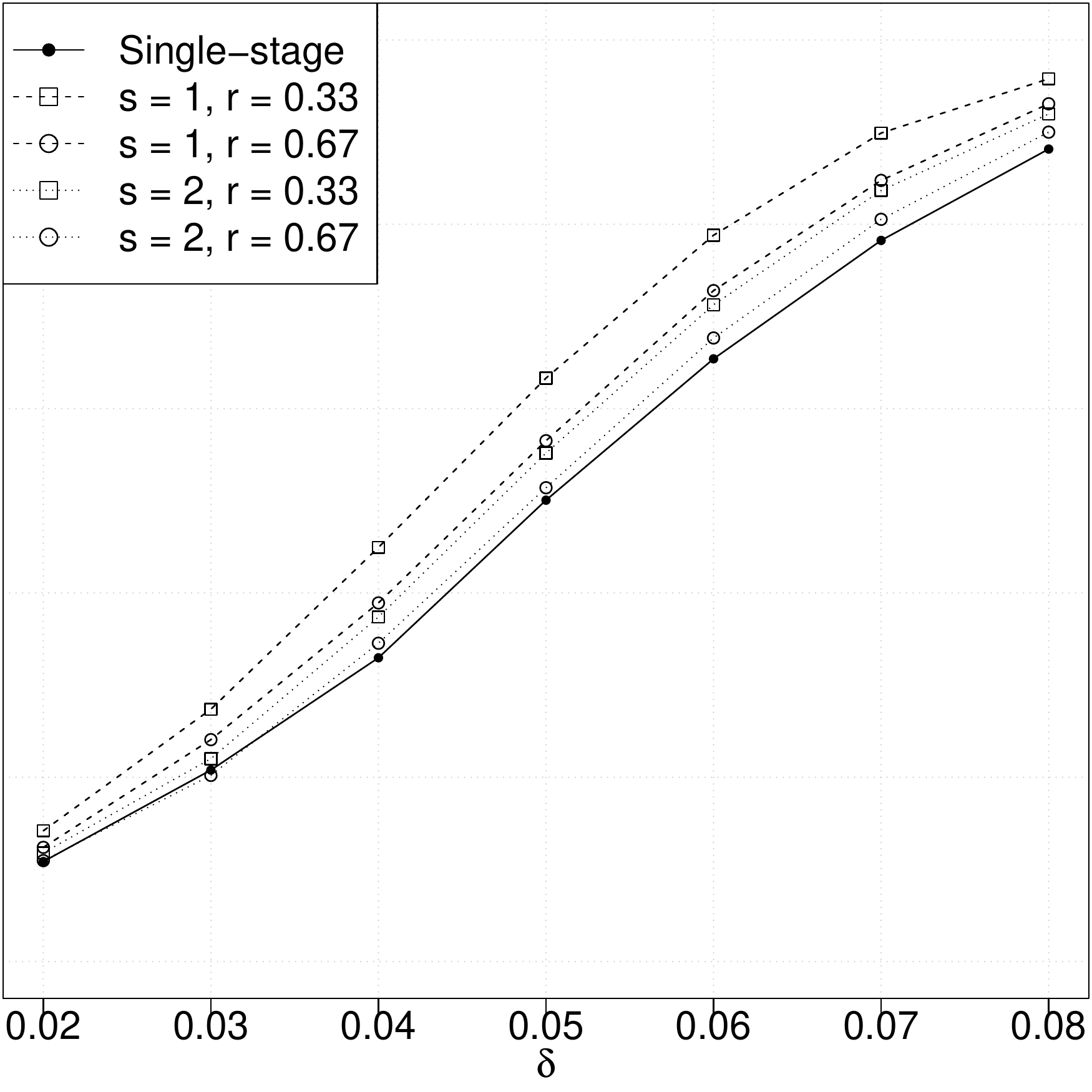} 
      \subcaption{\small{Effect sizes: $0$, $\delta - 2\eta$, $\delta - \eta$, $\delta$, with $\eta = 0.001$}}
      \label{fig:power_similar2}
  \end{subfigure}
  \caption{\small{Power to reject at least one hypothesis as a function of the effect size parameter $\delta$, for two different effect configurations given by Eqs.\ \eqref{eq:effect_config1}, \eqref{eq:effect_config2a} and \eqref{eq:effect_config2b}. Two different selection rules are compared. Either the single best treatment is taken forward to the second stage ($s = 1$), or the two best treatments are taken forward ($s = 2$). There are two options for the proportion of the total sample size used in the first stage ($r = 0.33$ or $r = 0.67$)}}
  \label{fig:power2}
\end{figure}

We next consider the case of a total sample size of $10{,}000$ and an expected effect size of $\delta^* = 0.05$. Moreover, there are now three active treatments in the first stage instead of just two as in Section \ref{sec:design_small_sample_large_effect}.\footnote{The two different cases have been chosen so as to roughly correspond to the two different applications. However, since the means of the Z-statistics have the form $\delta \sqrt{n / 2}$, the overall shape of the power curves is the same for the two cases. Therefore, from a purely mathematical viewpoint the two cases are actually very similar. Nevertheless, we choose to study them both using simulation in order to demonstrate the methodology.} In contrast to Figures \ref{fig:power_linear1} and \ref{fig:power_similar1}, it can be seen in Figures \ref{fig:power_linear2} and \ref{fig:power_similar2} that the single-stage design is dominated by all two-stage designs for all but the smallest effect sizes. This is because, when starting with three active treatments, at least one will always be dropped in the interim analysis. This leads to an improved power to reject at least one hypothesis.

It can be seen that the best design for both effect configurations is clearly the one corresponding to $s = 1$ and $r = 0.33$. As expected, for the case of similar effect sizes shown in Figure \ref{fig:power_similar2}, all power curves are shifted upwards since the effects are now all clustered close to $\delta$. Since it turns out that behaviour of the selection probabilities is very similar to that observed in Figures \ref{fig:selection_probs_linear1} and \ref{fig:selection_probs_similar1}, we do not include the corresponding plots here.\footnote{These plots can be reproduced by the open source code accompanying the paper.}  Again, we are led to a design which takes a single treatment forward ($s = 1$) and uses a ratio value that is somewhere between very small and $0.6$. 

\section{Applications} \label{sec:applications}

\subsection{Application 1: A business management game for regulatory impact analysis} \label{sec:musshoff_hirschauer} \label{sec:application1}

First, we apply a two-stage design inspired by the simulations in the previous section to a subset of the data from \citet{MusshoffHirschauer2014}. In their experiment, the impact of four different nitrogen reduction policies on the decision-making of the participants is investigated and compared to a base case. The participants play a multi-period, single-player business simulation game. The sample consists of data from 190 agricultural university students, taking on the role of farmers deciding on a production plan for a hypothetical crop farm. Each individual plan specifies the land usage shares for the different crops, as well as the amount of nitrogen fertiliser to be used. The objective of each player is to maximise his or her bank deposit at the end of the 20 production periods that make up the total game time.

In the first 10 production periods, all players receive an unconditional, deterministic payment. After these initial periods, the players are randomly assigned to one of five different treatment groups, representing different policy scenarios. The first scenario can be interpreted as ``business as usual'', in which the deterministic payments are continued. The other four are actual interventions, designed to foster more environmentally friendly production. They differ with respect to whether payments are granted (scenario 2 and 3), or whether penalties are charged (scenario 4 and 5) when the fertiliser usage is above, respectively below a given threshold (voluntary vs.\ prescriptive). Further, they vary in terms of whether the payments or penalties are made without exception (scenario 2 and 4), or whether there is only a certain probability that a player's action is discovered and payments or penalties are applied (scenario 3 and 5; deterministic vs.\ stochastic). The treatment designs ensure the same expected monetary gain for the players (see Section 3 of \citet{MusshoffHirschauer2014} for a detailed description of the experiment).

The outcome variable of main interest is the share of arable land where the players kept the fertiliser levels below a pre-specified threshold referred to as ``extensive use'', defined to be $75$ kg per hectare. In the first ten periods the group mean shares were similar, all between 11.6\% and 17.4\%. For periods 11 to 20 the shares were, for scenario 1 to 5 in order, 13.6\%, 67.5\%, 80.7\%, 80.9\% and 49.4\% (see Table 4 of \citet{MusshoffHirschauer2014}). Thus, scenarios 2-5 yield much higher share values compared to the reference scenario, and moreover these seem to lead to different outcome means although they all give the same expected profit.

Here, we will focus on the subset of data corresponding to scenarios 3, 4 and 2. For each of these groups, there is data for 38 players. As the total sample size increases in the examples below, more and more players are used from each group, in the order they are included in the original data file.\footnote{Table 4 of \citet{MusshoffHirschauer2014} can be reconstructed by taking the mean group values of the data used here (for maximum group sizes) and dividing by 400.} We use scenario 2 as the control (rather than scenario 1) and treat 3 and 4 as two different active treatments. We may interpret this as a smaller study in which a deterministic penalty (scenario 3) and a stochastic penalty (scenario 4) are compared against a deterministic reward (scenario 2). The null hypothesis for the comparison of scenario 3 vs.\ 2 is denoted by $H_1$, while the one corresponding to 4 vs.\ 2 is denoted by $H_2$. While scenario 2 (as a policy intervention) would be preferable based on equality arguments, high monitoring costs may make it impractical and warrant alternative interventions. Such are provided by scenarios 3 and 4, but which of them works best when taking into account the multiple goals (e.g.\ income and environment concerns) and bounded rationality of an economic decision maker?

Tables \ref{tab:musshoff_single_stage} and \ref{tab:musshoff_two_stage} provide a comparison showing which hypotheses are rejected for different subsets of the data. As in our simulation studies in Section \ref{sec:simulation}, groups within the same stage are always of equal size. Inspired by our simulation study in Section \ref{sec:simulation}, our selection rule at the interim analysis of the two-stage study takes the single best active treatment forward. Furthermore, approximately half the total sample size is used in the first stage. This design was chosen since it was found to be close to optimal in the simulations performed. It can be oserved that the two-stage design is able to reject at least one hypothesis (namely, $H_1$) once the total sample size $N$ becomes large enough, while the single-stage design never rejects any hypothesis. Hence, the adaptive design increases the power of detecting at least one effect by dropping one scenario at the interim analysis and focusing on the one that seems to have the largest effect. 

\begin{table}
  \centering
 
  \begin{subtable}{1.0\textwidth}
    \centering
    \begin{tabular}{ |l|c|l|l| } 
      \hline
      $n_1 \, (N_1)$ & $H_1 \cap H_2$ & $H_1 \, (\bar{x}_1 - \bar{x}_0)$ & $H_2 \, (\bar{x}_2 - \bar{x}_0)$ \\
      \hline
      10 (30) & \textcolor{red}{No} & \textcolor{red}{No} (8.00)  & \textcolor{red}{No} (39.30) \\
      13 (39) & \textcolor{red}{No} & \textcolor{red}{No} (35.38) & \textcolor{red}{No} (32.54) \\ 
      16 (48) & \textcolor{red}{No} & \textcolor{red}{No} (32.31) & \textcolor{red}{No} (25.63) \\ 
      20 (60) & \textcolor{red}{No} & \textcolor{red}{No} (36.65) & \textcolor{red}{No} (55.75) \\
      23 (69) & \textcolor{red}{No} & \textcolor{red}{No} (39.04) & \textcolor{red}{No} (50.22) \\
      26 (78) & \textcolor{red}{No} & \textcolor{red}{No} (41.38) & \textcolor{red}{No} (53.19) \\ 
      30 (90) & \textcolor{red}{No} & \textcolor{red}{No} (55.03) & \textcolor{red}{No} (52.60) \\ 
      \hline
    \end{tabular}
    \subcaption{\small{Single-stage design}}
    \label{tab:musshoff_single_stage}
  \end{subtable}
  
  \vspace{0.5cm}
  
  \begin{subtable}{1.0\textwidth}
    \centering    
    \begin{tabular}{ |l|l|c|l|l| } 
      \hline
      $n_1 \, (N_1) $ & $n_2 \, (N_2)$ & $H_1 \cap H_2$ & $H_1 \, (\bar{x}_1 - \bar{x}_0)$ & $H_2 \, (\bar{x}_2 - \bar{x}_0)$ \\
      \hline
      5  (15) &  7 (14) & \textcolor{red}{No} & \textcolor{red}{No} (12.83) & \textcolor{red}{No} (64.42) \\
      7  (21) &  9 (18) & \textcolor{red}{No} & \textcolor{red}{No} (24.28) & \textcolor{red}{No} (25.63) \\ 
      9  (27) & 11 (22) & \textcolor{red}{No} & \textcolor{red}{No} (30.77) & \textcolor{red}{No} (55.75) \\ 
      10 (30) & 15 (30) & \textcolor{red}{No} & \textcolor{red}{No} (39.00) & \textcolor{red}{No} (53.72) \\ 
      12 (36) & 17 (34) & \textcolor{red}{No} & \textcolor{red}{No} (22.24) & \textcolor{red}{No} (49.07) \\ 
      14 (42) & 19 (38) & \textcolor{green}{Yes} & \textcolor{green}{Yes} (60.67) & \textcolor{red}{No} (29.51) \\ 
      15 (45) & 22 (44) & \textcolor{green}{Yes} & \textcolor{green}{Yes} (58.57) & \textcolor{red}{No} (36.00) \\ 
      \hline
    \end{tabular}
    \subcaption{Two-stage design}
    \label{tab:musshoff_two_stage}
  \end{subtable}
  
  \caption{\small{Hypotheses rejected in Dunnett tests for single- and two-stage designs of increasing sizes, together with mean difference estimates of effect sizes. Note that the total sample sizes for the corresponding rows of Tables \ref{tab:musshoff_single_stage} and \ref{tab:musshoff_two_stage} have been selected to be as close as possible. Type I error probability $\alpha = 0.025$. $n_1$ = first stage group size, $n_2 = $ second stage group size. $N_1 = $ total first stage sample size, $N_2 = $ total second stage sample size}}
  \label{tab:musshoff}
\end{table}

\subsection{Application 2: A field experiment in charitable giving behaviour} \label{sec:application2}

Next, we consider the study of \citet{KarlanList2007}, which is concerned with the effect of a matching grant on charitable giving. A matching grant represents an additional donation made by a central donor backing the study, the size of which depends on the amount donated by the regular donors. The sample consists of data from 50{,}083 individuals, which were identified as previous donors to a non-profit organisation and were contacted by mail.

Two thirds of the study participants were assigned to some active treatment group, while the rest were assigned to a control group. The active treatments varied along three dimensions: (1) the price ratio of the match (\$1:\$1, \$2:\$1 or \$3:\$1)\footnote{The ratio \$X:\$1 indicates that for every dollar the individual donates, the matching donor additionally contributes \$X.}; (2) the maximum size of the matching gift based on all donations (\$25,000, \$50,000, \$100,000 or unstated) and (3) the donation amount suggested in the letter (1.00, 1.25 or 1.50 times the individual’s highest previous contribution). Each of these treatment combinations was assigned with equal probability. The authors study both the effects on the response rate as well as the donation amount. However, we will only focus on how the matching ratio affects the response rate. The reason for this simplification is that we want to be able to connect the simulation results in the previous section to the available real-world data without increasing the complexity of the presentation by introducing too many treatment groups.

For illustrative purposes, we will only consider a portion of the data analysed by \citet{KarlanList2007}. Specifically, we will restrict ourselves to the subset consisting of those individuals residing in red states (i.e., states with a republican majority, see Panel C of Table 2A). The reason for this is that, as noted by \citet{KarlanList2007}, it is only for this subgroup that the original analysis shows a significant difference between the groups defined by the different matching ratios. Naturally, a real analysis would not ignore data just because it is not significant. However, since our purpose here is to compare methods of data analysis rather than to answer specific research questions, we choose to restrict ourselves to a subset in order to be able to more clearly highlight the difference between the single-stage and two-stage designs. Thus, there are three active treatments, corresponding to the different matching ratios. As in the simulations and the previous application, the sample size is divided equally among the groups in each stage. We again employ a rule that selects the single best treatment at the interim analysis and use a ratio value of 0.5.

\begin{table}
  \centering
  \begin{subtable}{1.0\textwidth}
    \centering
    \begin{tabular}{ |l|c|l|l|l| } 
      \hline
      $n_1 \, (N_1)$ & $H_1 \cap H_2 \cap H_3$ & $H_1 \, (\bar{x}_1 - \bar{x}_0)$ & $H_2 \, (\bar{x}_2 - \bar{x}_0)$ & $H_3 \, (\bar{x}_3 - \bar{x}_0)$ \\
      \hline
      500   (2,000)  & \textcolor{red}{No}    & \textcolor{red}{No}    (0.004) & \textcolor{red}{No}    (-0.006) & \textcolor{red}{No}    (0.002) \\
      750   (3,000)  & \textcolor{red}{No}    & \textcolor{red}{No}    (0.008) & \textcolor{red}{No}    (-0.005) & \textcolor{red}{No}    (0.000) \\ 
      1,000 (4,000)  & \textcolor{red}{No}    & \textcolor{red}{No}    (0.009) & \textcolor{red}{No}    (0.000)  & \textcolor{red}{No}    (0.006) \\ 
      1,250 (5,000)  & \textcolor{red}{No}    & \textcolor{red}{No}    (0.010) & \textcolor{red}{No}    (0.002)  & \textcolor{red}{No}    (0.005) \\
      1,500 (6,000)  & \textcolor{red}{No}    & \textcolor{red}{No}    (0.011) & \textcolor{red}{No}    (0.006)  & \textcolor{red}{No}    (0.007) \\
      1,750 (7,000)  & \textcolor{green}{Yes} & \textcolor{green}{Yes} (0.014) & \textcolor{red}{No}    (0.008)  & \textcolor{red}{No}    (0.010) \\ 
      2,000 (8,000)  & \textcolor{green}{Yes} & \textcolor{green}{Yes} (0.012) & \textcolor{red}{No}    (0.008)  & \textcolor{green}{Yes} (0.010) \\
      2,250 (9,000)  & \textcolor{green}{Yes} & \textcolor{green}{Yes} (0.010) & \textcolor{red}{No}    (0.008)  & \textcolor{green}{Yes} (0.012) \\
      2,500 (10,000) & \textcolor{green}{Yes} & \textcolor{green}{Yes} (0.012) & \textcolor{green}{Yes} (0.009)  & \textcolor{green}{Yes} (0.014) \\
      2,750 (11,000) & \textcolor{green}{Yes} & \textcolor{green}{Yes} (0.013) & \textcolor{green}{Yes} (0.009)  & \textcolor{green}{Yes} (0.014) \\ 
      \hline
    \end{tabular}
    \subcaption{\small{Single-stage design}}
    \label{tab:karlan_list_single_stage}
  \end{subtable}
  
  \vspace{0.5cm}
  
  \begin{subtable}{1.0\textwidth}
    \centering    
    \begin{tabular}{ |l|l|c|l|l|l| } 
      \hline
      $n_1 \, (N_1)$ & $n_2 \, (N_2)$ & $H_1 \cap H_2 \cap H_3$ & $H_1 \, (\bar{x}_1 - \bar{x}_0)$ & $H_2 \, (\bar{x}_2 - \bar{x}_0)$ & $H_3 \, (\bar{x}_3 - \bar{x}_0)$ \\
      \hline
      250   (1,000) & 500   (1,000) & \textcolor{red}{No}    & \textcolor{red}{No}    (0.008) & \textcolor{red}{No} (0.007) & \textcolor{red}{No} (0.003) \\
      375   (1,500) & 750   (1,500) & \textcolor{red}{No}    & \textcolor{red}{No}    (0.009) & \textcolor{red}{No} (0.006) & \textcolor{red}{No} (0.012) \\ 
      500   (2,000) & 1,000 (2,000) & \textcolor{green}{Yes} & \textcolor{green}{Yes} (0.011) & \textcolor{red}{No} (0.005) & \textcolor{red}{No} (0.013) \\ 
      625   (2,500) & 1,250 (2,500) & \textcolor{green}{Yes} & \textcolor{green}{Yes} (0.013) & \textcolor{red}{No} (0.005) & \textcolor{red}{No} (0.013) \\ 
      750   (3,000) & 1,500 (3,000) & \textcolor{green}{Yes} & \textcolor{green}{Yes} (0.010) & \textcolor{red}{No} (0.004) & \textcolor{red}{No} (0.009) \\ 
      875   (3,500) & 1,750 (3,500) & \textcolor{green}{Yes} & \textcolor{green}{Yes} (0.012) & \textcolor{red}{No} (0.006) & \textcolor{red}{No} (0.008) \\ 
      1,000 (4,000) & 2,000 (4,000) & \textcolor{green}{Yes} & \textcolor{green}{Yes} (0.010) & \textcolor{red}{No} (0.003) & \textcolor{red}{No} (0.009) \\
      1,125 (4,500) & 2,250 (4,500) & \textcolor{green}{Yes} & \textcolor{green}{Yes} (0.008) & \textcolor{red}{No} (0.002) & \textcolor{red}{No} (0.005) \\
      1,250 (5,000) & 2,500 (5,000) & \textcolor{red}{No}    & \textcolor{red}{No}    (0.006) & \textcolor{red}{No} (0.001) & \textcolor{red}{No} (0.004) \\
      1,375 (5,500) & 2,750 (5,500) & \textcolor{red}{No}    & \textcolor{red}{No}    (0.006) & \textcolor{red}{No} (0.003) & \textcolor{red}{No} (0.003) \\
      \hline
    \end{tabular}
    \subcaption{Two-stage design}
    \label{tab:karlan_list_two_stage}
  \end{subtable}
  \caption{\small{Hypotheses rejected in Dunnett tests for single- and two-stage designs of increasing sizes, together with mean difference estimates of effect sizes. Note that the total sample sizes for the corresponding rows of Tables \ref{tab:karlan_list_single_stage} and \ref{tab:karlan_list_two_stage} have been selected to be as close as possible. Type I error probability $\alpha = 0.025$. $n_1$ = first stage group size, $n_2 = $ second stage group size. $N_1 = $ total first stage sample size, $N_2 = $ total second stage sample size}}
  \label{tab:karlan_list}
\end{table}

\begin{figure}[!ht]
  \centering
  \begin{subfigure}[b]{0.49\textwidth}
    \centering
    \includegraphics[width=\textwidth]{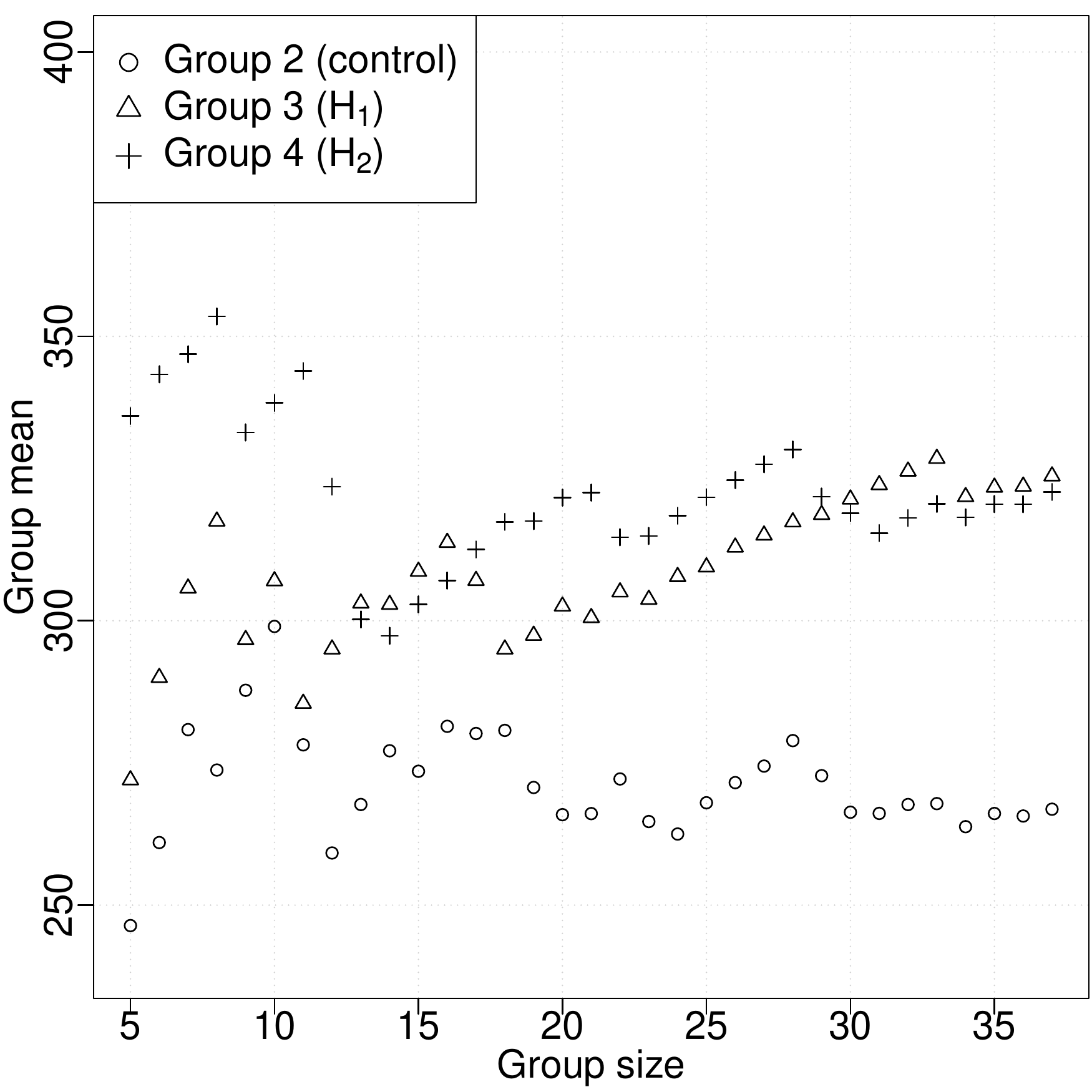} 
    \subcaption{\small{Application 1}}
    \label{fig:group_means1}
  \end{subfigure}
  \hfill
  \begin{subfigure}[b]{0.49\textwidth}
    \centering
      \includegraphics[width=\textwidth]{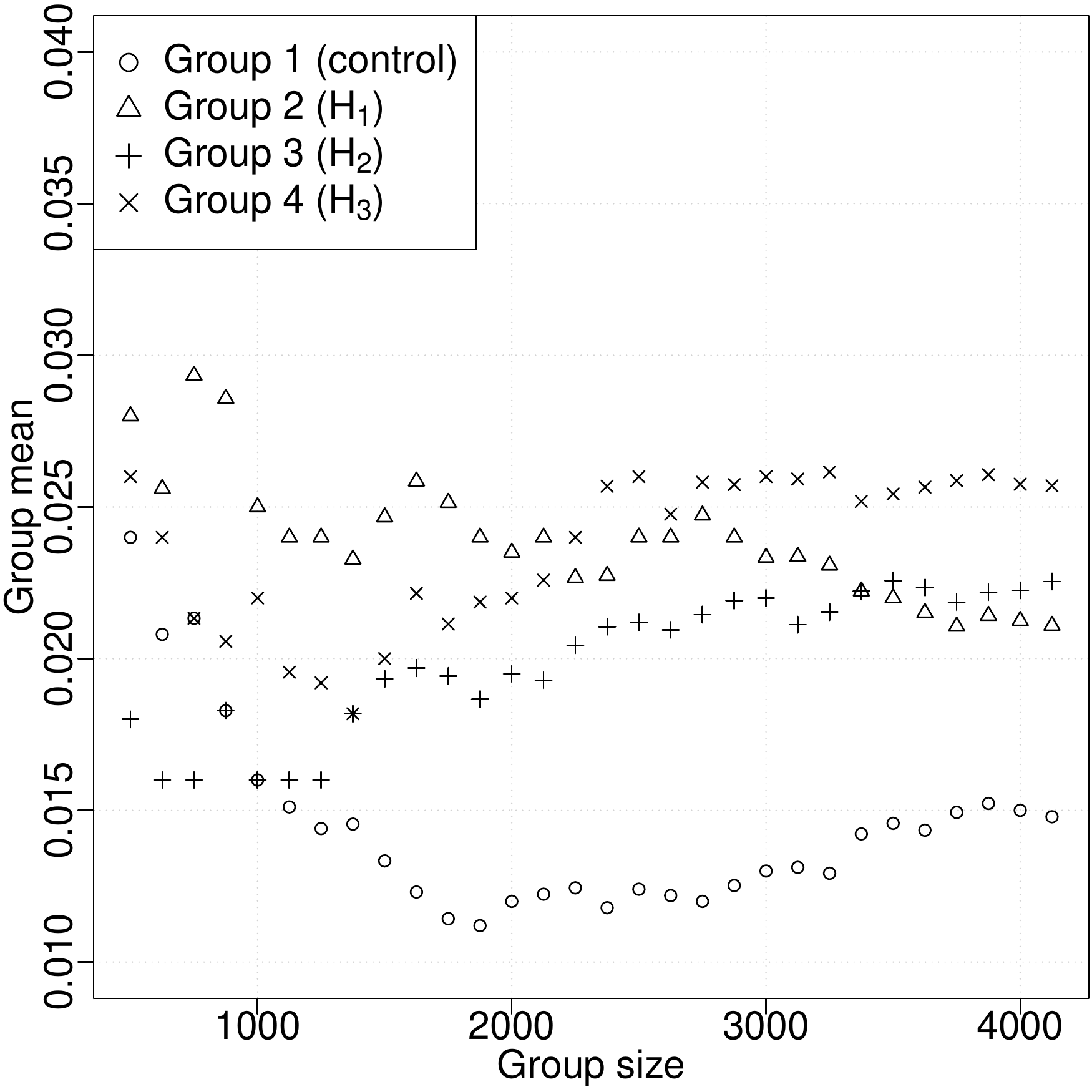} 
      \subcaption{\small{Application 2}}
      \label{fig:group_means2}
  \end{subfigure}
  \caption{\small{Goup means as functions of group sample sizes}}
  \label{fig:group_means}
\end{figure}

Comparing Table \ref{tab:karlan_list_single_stage} with Table \ref{tab:karlan_list_two_stage}, it can be seen that the single-stage design rejects no hypothesis up until a total sample size of $N_1 = 6{,}000$. In contrast, the two-stage design starts to reject a hypothesis, namely $H_1$, from $N_1 + N_2 = 4{,}000$. Interestingly, this pattern of rejection stops at $N_1 + N_2 = 10{,}000$ and the test is no longer able to reject any hypothesis for the two largest values of $N_1$ and $N_2$. In order to explain why this non-monotonic behaviour occurs, a more detailed examination of the data is needed. Figure \ref{fig:group_means2} shows the group means for the different groups as functions of the increasing group sample sizes. When the rejection starts at the third row in Table \ref{tab:karlan_list_two_stage}, the decision to select group 2 for the second stage is based on the means of the first $500$ subjects from each group. Since group 2 has the largest mean, it is selected and taken forward to the second stage. After the second stage, a total of $1{,}500$ subjects from group 2 (and the control group) are used to form the means, and the mean difference is now even larger than after the first stage. At the next to last row of Table \ref{tab:karlan_list_two_stage}, when the total sample size is $N_1 + N_2 = 10{,}000$, none of the hypotheses are rejected anymore. When the reversal occurs, $3{,}750$ subjects are used from group 2. However, at this group size group 2 is no longer the best active group, but the worst one, since a change in order happens at a group size of $2{,}250$, after which group 3 becomes the best one. The lower group mean value of group 2 is enough to destroy the significant result. The single-stage design, on the other hand, starts to reject all hypotheses once the total sample size becomes large enough and does not show the same reversal. The reason is that the latter part of the data contributing to the smaller mean values of group 2 are not included. This reversal can be compared to the behaviour observed for the first application in Figure \ref{fig:group_means1}. There, the group means of the two active groups converge to a smiliar value as the group sizes increase, and there is no apparent reversal. 

There are multiple potential explanations for the behaviour of the mean of group 2 observed in Figure \ref{fig:group_means2}. For example, assuming that data entries are ordered according to reponse date, the donations may have come from lower income quantiles facing liquidity constraints at the time the donation call was received, leading to delayed responses. As discussed by the authors of the original study, turnaround time (the time required for shipping the donation call and shipping time for the donation checks) may differ by weeks in nationwide studies in large countries like the USA, leading to time-dependent patterns in the data. Taking into account further differences between deliveries in urban and rural areas, it appears likely that responses from a) more distant states and b) more rural areas, are reported later. The important point we want to stress here is that if there are reasons to suspect a time trend in the data, then the researcher must be very careful when applying adaptive designs.

\section{Discussion and conclusions} \label{sec:discussion}
In this paper, we have introduced the basic theory of adaptive designs for multiple hypothesis testing and illustrated it by means of two real-world data sets from experimental economics. Although the basic theory required for the applications is covered by Section \ref{sec:theory}, the general approach of using combination tests and the closed testing principle provides a starting point for more advanced designs. Using the R package \texttt{asd} for our simulation study reflects our aim to introduce and illustrate the available methodology, since it is freely available and well suited to the scale of our applications. We chose to include two studies since they differ in distinct dimensions which are relevant for economic experiments. First, they represent different experiment types (natural field experiment vs. framed field experiment; cf. \citet{harrison_field_2004}). Although the experiment described by \citet{MusshoffHirschauer2014} ensured incentive compatibility, the experiment contained hypothetical decision making, whereas the donation decisions in \citet{KarlanList2007} were non-hypothetical. Moreover, the study of \citet{MusshoffHirschauer2014} analyses data from a so-called convenience group. Second, the experiments differ with respect to their complexity (participants having only to decide upon whether to donate and the respective amount vs. participants having to play a complex multi-period business simulation game). Finally, the studies vary with respect to the sample sizes as well as the reported effect sizes. This allows us to illustrate the general usefulness, and potential pitfalls, of adaptive designs for applications in experimental economics.

Based on the results presented in the previous section, the question remains which gains that can be expected by moving from a single-stage to a two-stage design. To get an overall idea about the kind of improvements to expect, we can return to our simulation results. Consider Figures \ref{fig:power_linear1}, \ref{fig:power_similar1}, \ref{fig:power_linear2} and \ref{fig:power_similar2}. Starting with the case of a small sample and a large effect in Figure \ref{fig:power_linear1}, it can be seen that the best two-stage design considered ($s = 1$, $r = 0.33$) always has a higher power to reject at least one hypothesis. The largest gain, about 7 percentage points, is obtained around $\delta = 0.6$. A similar gain is obtained for the case of almost equal effects sizes shown in Figure \ref{fig:power_linear2}, over the region $\delta \in [0.4, 0.6]$. For the other case of a large sample and a small effect, the gains are larger. They reach about 15 percentage points in the region $\delta \in [0.4, 0.6]$ for linearly distributed effect sizes in Figure \ref{fig:power_linear2}, and are of a similar magnitude for the case of similar effect sizes in Figure \ref{fig:power_similar2}. Clearly, the observation on the possible gain in power made here is very specific. However, it can still serve as an approximate estimate of the gain that we can expect when moving from a single-stage to a two-stage design. While these results show the potential of applying two-stage designs, the decision to apply these require additional practical considerations. Most importantly, the costs associated with the administration and data analysis of the more complex two-stage design compared with a single-stage design have to be taken into account. For example, when field experiments are carried out in the paper-and-pencil-format, digitising and analysing the interim data may significantly increase the study's costs (or even be logistically infeasible). Still, when the data is sequentially collected in digital form (e.g. during multiple experimental sessions in an economic laboratory) a two-stage design may not impose any additional costs at all.

The relatively simple models considered in the present paper can be extended in a number of different ways. One option would be to consider designs with more than two stages. Whether or not this is worthwhile from a practical perspective depends to a high degree on the specific application considered. Again, if experimental data are sequentially collected in multiple sessions and in a form which can be analysed without technical difficulties, multi-stage designs could be a viable option. Here, it is important to emphasise that this will not pose any issues from a theoretical perspective, since the general method of using combination tests and the closed testing principle would still be applicable. Nevertheless, the simulations would require additional programming since the \texttt{asd} package only supports two stages. Alternatively, other (mainly commercial) software packages are available which could be considered when implementing more advanced analyses.

It is further important to note that alternative selection rules could have been considered in the previous sections. To simplify the presentation of our simulation results, we always used rules that took a fixed number of treatments forward to the next stage. But the general framework of adaptive designs allows for essentially \emph{any} type of rule. For example, we could have selected all the treatments with effect estimates within a certain range of the best one. This is the so called $\epsilon$-rule and is supported by the \texttt{asd} package. Similarly, while the use of Dunnett's test for many-to-one comparisons fits our two applications, it is but one of several possible choices for the intersection test to use when applying the closed testing principle. Again, we want to stress that our choice here was for illustrative purposes and that alternative applications may require other types of tests. Furthermore, when a researcher wants to carry out large-scale experiments with a very large number of hypotheses tested simultaneously, strong control of the FWER may not be an appropriate criterion. A possible alternative that has been suggested for such cases is the per comparison error rate mentioned in Section \ref{sec:theory}. An example from medicine for such large-scale multiple testing is the measurement of the expression levels of thousands of genes simultaneously using microarrays.

In our analysis we focused exclusively on hypothesis testing as the main goal of the experiment. However, formulating and testing null hypotheses is often only partly the goal of similar studies. In addition, the experimenter often seeks to perform some kind of model or parameter estimation. In medicine, this could consist of reporting estimates and confidence intervals for the different treatments considered. When constructing estimates from the data obtained in an adaptive experiment, it is important to note that these may be biased. Just applying the fixed-design methods for computing point estimates or confidence intervals is typically not enough, since special adjustments based on the adaptions performed are required. A simple example of the complications that might arise is provided by \citet{Pallmann2018}. Consider a trial with one interim analysis comparing a single active treatment to control, and suppose that the adaptive rule consists of stopping the trial for futility if the effect estimate at the interim is small enough. Suppose further that the effect estimate for non-stopping trials is computed by combining all data from both stages. This combined estimate will then be biased upwards, since effect estimates that are lower simply by chance leads to stopping. Although we recognise the importance of estimation problems, our aim here has been to keep the complications at a minimum and focus on the issue of strong FWER control in the presence of interim analyses.
  
To connect this general issue to one of our applications, consider the analysis by \citet{MusshoffHirschauer2014}. There, the objective is not only to determine whether a specific set of hypotheses can be rejected or not. The results are also used to fit a regression model. While the confidence intervals obtained for the estimated parameters are used to check the hypotheses, the value of such a model goes beyond testing since it can also be used for more general predictive statements. The apparent time trend observed in the data from \citet{KarlanList2007} highlights another issue. When the treatment effects changes between the different stages of the experiment, the conclusions drawn from the interim analysis may become biased or unreliable. While this problem could have been solved in the presented application by randomising the order of the data, it serves as a warning that in actual applications a cautious evaluation of potential (hidden) pitfalls in the data collection is required. If the design cannot rule out systematic heterogeneity between the experimental stages, researchers may also address these issues by formal testing procedures (see e.g. \citet{friede2009exploring}).

The \texttt{asd} package supports a number of related design approaches, which could be considered for economic experiments. Instead of the many-to-one comparisons that have been our focus here, similar methods can also be applied to the closely related topic of subgroup selection, in which one or more subgroups of treatment units are taken forward to the next stage in a sequential study while the treatment stays the same. These are referred to as \emph{adaptive enrichment designs} (see, e.g., Chapter 11 of \citet{WassmerBrannath2016}). Simulations for a special case of such designs, involving co-primary analyses in a pre-defined subgroup and the full population, are supported by the \texttt{asd} software. Another feature supported by the \texttt{asd} package which we did not employ is that of simulating trials with \emph{early outcomes}. This means that the decision concerning which treatment to take forward to the second stage is based on a measure that need not be identical but only correlated to the one actually used in the final analysis. Early outcomes are important in medicine, where final trial results may take months or even years to collect, but where it is still valuable to be able to use some correlated measure to steer treatment allocation. As a relevant example from economics, we may take the study by \citet{MusshoffHirschauer2014}. While the data of that study is provided by the participants of a business simulation game, the real objective is of course to evaluate how the different incentive structures considered impact the decision making of real farmers. One could then view the decisions of the game participants as early outcomes in a wider study that would evaluate some subset of selected incentives using data obtained from real farmers. 

An approach that is closely connected to general adaptive designs is that of \emph{group-sequential designs}. For these, a maximum number of groups is first pre-specified. Subjects are then recruited sequentially and an interim analysis is performed once the results for a full group have been obtained. The value of a summary statistic that includes the information from all groups seen so far is computed and used to decide whether to stop or continue the trial. The main body of the theory is built on the often reasonable assumption of a normally distributed summary statistic. It turns out that there is then a special technique, the \emph{recursive integration formula} \citep{Armitage1969}, which can be used to compute the characteristics in a numerically efficient manner. A number of different group-sequential designs have been suggested in the literature (\citet{Pocock1977}, \citet{OBrienFleming1979}, etc.). These classical designs can be seen as special cases of the more general adaptive approach that we have used in this paper. Whereas the group-sequential approach may lead to a certain degree of flexibility in the total sample size of the trial by allowing for early termination at an interim analysis, it does not allow for things such as sample size reassessment or the dropping of one or several treatment arms. For a comprehensive treatment of the group-sequential approach, see for example \citet{JennisonTurnbull2000}.

Summarising, the general framework outlined in the present paper is a valuable extension of the methodological toolkit in experimental economic research. It can be applied in many experimental settings using existing, open-source software and has the potential to foster the efficient usage of the resources available to the researcher.


\section*{Conflict of interest}
The authors declare that they have no conflict of interest.

\bibliographystyle{apalike} 
\bibliography{jobjornsson_schaak_musshoff_friede_2021}   

\appendix

\section{Experimental Instructions} \label{sec:appendix}

This appendix contains a more detailed description of the experiments serving as our two applications, and the simulation study presented in Section \ref{sec:simulation}.

\subsection{Application 1} \label{sec:appendix1}
In this application, by \citet{MusshoffHirschauer2014}, there were five different treatment groups, referred to as scenarios, and 38 students were randomised to each scenario group. The participants were on average 25 years old, and 38\% were female. There are three different kinds of crops: wheat, canola and potatoes. Each individual plan specifies the land usage shares for the different crops, as well as the amount of nitrogen fertiliser to be used. This is done under the constraints that a total of 400 ha are available for cultivation, and that each crop must be assigned at least 100 ha. The objective of each player is to maximise his or her bank deposit at the end of the 20 production periods that make up the total game time. The only sources of risk are the market prices, which change over time as stochastic processes.

Each player is provided with a starting capital of 1 million euros. At the beginning of each period, \euro 40{,}000 are withdrawn for personal living expanses. No interest is paid on positive credit balance at the end of a period. If there are outstanding payments that the player is unable to settle, capital is borrowed at zero interest rate. This is then automatically repaid in later periods if liquid funds then exceed the obligatory withdrawal. Fieldwork is assumed to be carried out by contractors at crop specific cost rates.

The production functions for the three different crops are all of the form
\begin{equation}
  y = a + bx - c x^2 , 
\end{equation}
where $x$ is the amount of nitrogen used (kg/ha) and $y$ is the yield (dt/ha, decitonnes per hectare). The parameters $a$, $b$ and $c$ are specific for each crop and participant, determined by drawing from a uniform distribution. The nitrogen price is randomly determined for each participant, and stays fixed during that person's game. At the end of each period, a specific crop price either increases to 120\% of what it was in the previous period or decreases to 80\% (each with probability 0.5). By design, no correlation exists between product prices. Start prices, in \euro/dt, were 20 for wheat, 30 for canola and 10 for potatoes. To incentivise the players, they are informed that 5\% of them (randomly selected) will receive a cash prize, the size of which depends on the performance in the game. \euro 500 was given to players who achieved the best possible bank balance, while less successful players were awarded a share of the prize according to their relative performance.

In the first 10 production periods, all players receive an unconditional, deterministic payment of \euro 300/ha. After these initial periods, the players are randomly assigned to one of five different treatment groups, representing different policy scenarios. The first, scenario 1, can be interpreted as ``business as usual'', in which the deterministic payments are continued. The other four are actual interventions, designed to foster more environmentally friendly production. Scenario 2 is a voluntary scheme in which players receive \euro 300/ha (for each ha individually) if nitrogen levels are kept below 75 kg/ha. Complete visibility of nitrogen use is achieved through a monitoring system. In scenario 3, the transfer payments of \euro 300/ha continues, but players will have to pay a penalty of \euro 300/ha for every hectare where the nitrogen limit of 75 kg/ha is exceeded. Again, there is full visibility due to a monitoring system. Scenario 4 is like scenario 3, except that now compliance to the nitrogen limit is investigated in random sample inspections. The probability to be investigated is 0.1, and upon detection of nitrogen use above the 75 kg/ha limit, a penalty of \euro 3000/ha has to be paid. Finally, in scenario 5, the players receive a transfer of \euro 3000/ha for those areas where the nitrogen use is below the 75 kg/ha limit, but only if they are inspected, which happens with probability 0.1. These different scenarios all ensure the same expected monetary gain for the players, provided that they play optimally. To facilitate comparability, the participants are divided into quintets. Each of the five groups includes one participant who is confronted with the same production
functions, the same nitrogen price and the same product price paths as the other members of the quintet in the other scenarios.

\subsection{Application 2} \label{sec:appendix2}
The second example \citep{KarlanList2007} used a natural field experiment to study the effects of different matching grants on charitable donations. The experiment targeted previous donors of a non-profit organisation and is based on a large sample of 50{,}083 individuals. Two thirds of the study participants were assigned to some active treatment group, while the rest were assigned to a control group. The active treatments varied along three dimensions: (1) the price ratio of the match (\$1:\$1, \$2:\$1 or \$3:\$1); (2) the maximum size of the matching gift based on all donations (\$25,000, \$50,000, \$100,000 or unstated) and (3) the donation amount suggested in the letter (1.00, 1.25 or 1.50 times the individual’s highest previous contribution). Each of these treatment combinations was assigned with equal probability.

Ask amounts were specific for each participant and based on the individual's highest previous contribution. To test for additional effects, such as solicitee income level and political affiliation, the giving data was merged with: (1) demographic data aggregated at zip code level, (2) state and county returns from the 2004 presidential election, and (3) frequency of activities of the organisation asking for donations at the state level. The main response variables of interest for us were the binary decision of whether or not to donate, and the property of belonging to a red state (republican voter).

The original data set used by \citep{KarlanList2007} was downloaded from an online repository. The main data file, \texttt{AER merged.dta}, was converted to an R file and then used to produce the plots and tables included in our paper. When loaded in R, the table corresponding to \texttt{AER merged.dta} contains a column \texttt{red0} which contains 1 at a specific row if it corresponds to a respondent from a red state (republican majority). As mentioned above in the main text, we only considered the subset of the data for which \texttt{red0 = 1}.

\subsection{Simulation study} \label{sec:appendix3}
In this section we will consider the simulation study described in Section \ref{sec:simulation} in more detail.

\subsubsection{General simulation procedure}
Based on the assumption of a fixed total budget, all the designs compared are assumed to have the same total sample size. Ignoring any additional complexity costs associated with two-stage vs.\ single-stage designs, this means that all designs have the same fixed cost. Hence, they may be compared in terms of their power, without any need to relate the power to the cost.

For the two-stage designs, we have assumed a specific class of selection rules, namely, the ones that always takes a fixed number of treatments forward to the second stage. The number of treatments taken forward is specified by the parameter $s$. This selection rule works by comparing the Z-values corresponding to the different stage 1 treatments relative to control and selecting the $s$ ones with the largest (absolute) Z-values. Hence, $s = 1$ means that the single best treatment is taken forward, $s = 2$ that the two best are taken forward, and so on. The second parameter characterising the two-stage adaptive designs is the ratio $r$, defined as the proportion of the total sample size used in the first stage. We now give a slightly more detailed description than the one provided in Section \ref{sec:simulation} of the procedure used to decide on the values of these parameters:
\begin{enumerate}
\item Specify a total sample size for the trial, and an \emph{expected effect size} $\delta^*$. The parameter $\delta^*$ allows for considering studies with different effect strengths. Typically, higher values of $\delta^*$ means that it will be easier to reach the primary goal and reject at least one hypothesis.
  
\item Plot the power to reject at least one hypothesis as a function of effect size, for different combinations of the selection rule $s$ and the ratio $r$. Choose the design which corresponds to the highest power in a neighbourhood of $\delta^*$. We refrain from specifying exactly how large this neighbourhood is, but some fraction of $\delta^*$ seems reasonable. The idea is that we want to choose a design that is reasonably robust with respect to uncertanties in our knowledge of the true effect sizes of the different treatments. 

\item For the specified value of $\delta^*$, fix the selection rule and ratio found in the previous step and plot the power and selection probabilities in a neighbourhood of the ratio. Adjust the ratio slightly if it leads to better selection probabilities without sacrificing too much power, say at most 1 or 2 percentage points. The idea behind this last step is to improve the performance in terms of a secondary criterion (selection probabilities) without sacrificing too much with respect to our primary criterion (power to reject at least one hypothesis). Of course, in a full formal optimisation problem, the relative importance of these two criteria would have to be specified in detail.
\end{enumerate}

The power in our simulations was always evaluated relative to one of two specific configurations of the treatment effects. In the first, the effects are increasing from $0$ to $\delta$ in steps of equal length, 
\begin{equation} 
  \mu_i = i \delta / m, \quad i = 0, \ldots, m.
\end{equation}
In the second configuration, the effect sizes of the active treatments are instead clustered close to $\delta$,
\begin{align} 
  \mu_0 & = 0, \\ 
  \mu_i & = \delta - (m - i) \eta, \quad i = 1, \ldots, m, 
\end{align}
for some small fixed number $\eta$. Naturally, we could also consider other types of effect configurations. In a fully specified optimisation problem, we could have a finite number of such configurations, together with an associated prior distribution over them specifying how likely we consider each of them to be a priori. 

\subsubsection{R code supporting the simulation study}
Instead of expanding on our treatment of the specific simulation results presented in Section \ref{sec:simulation}, we will now briefly consider the structure of the R code used to perform the simulations, and how this R code makes use of the package \texttt{asd}. We begin by stressing that the code we have implemented on top of \texttt{asd} is very specific, and essentially only contains the minimum functionality required to reproduce the computations and figures included in the present paper. It should be useful for those new to the adaptive design approach and interested in applying the methods implemented by \texttt{asd} to their own design problems.

The code consists of a single R source file, called \texttt{design\_opt\_submission.R}. It makes use of two R packages, \texttt{asd} and \texttt{mvtnorm}, which needs to be installed before running the functions in the file. The code essentially consists of a number of individual functions that employs the interface provided by the \texttt{asd} package to produce the figures and tables included in the paper. Specifically, Table \ref{tab:fun_descs} provides an overview of the central top-level functions in the file and the corresponding figures.

\begin{table}[!h]
  \centering 
  \begin{tabular}{|l|l|} 
    \hline
    Figure & Function \\
    \hline
    Figure \ref{fig:power_linear1} & \texttt{plot.power.linear1}  \\
    Figure \ref{fig:power_similar1} & \texttt{plot.power.similar1} \\
    Figure \ref{fig:selection_probs_linear1} & \texttt{plot.power.selection.probs.linear1} \\
    Figure \ref{fig:selection_probs_similar1} & \texttt{plot.power.selection.probs.similar1} \\
    Figure \ref{fig:power_linear2} & \texttt{plot.power.linear2} \\
    Figure \ref{fig:power_similar2} & \texttt{plot.power.similar2}  \\
    Figure \ref{fig:group_means1} & \texttt{plot.Musshoff.Hirschauer.group.means} \\
    Figure \ref{fig:group_means2} & \texttt{plot.Karlan.List.group.means} \\    
    \hline
  \end{tabular}  
  \caption{\small{Functions implemented in the file \texttt{design\_opt\_submission.R}}}
  \label{tab:fun_descs}
\end{table}

The functions \texttt{plot.power.linear1}, \texttt{plot.power.similar1}, \texttt{plot.power.linear2} and \texttt{plot.power.similar2} are all implemented as direct calls to the function \texttt{plot.power}, which allows for specifying the details of the simulation, such as the available values of $s$ and $r$, the significance level, and the effect configuration. In turn, \texttt{plot.power} performs the major simulation work by calling the functions \texttt{single.stage.sim} and \texttt{two.stage.sim}, both of which are implemented via the iterface provided by the \texttt{asd} package. The functions \texttt{plot.power.selection.probs.linear1} and \texttt{plot.power.selection.probs.similar1} are implemented by calling \texttt{plot.power.selection.probs}, which in turn essentially consists of repeated calls to the already mentioned \texttt{two.stage.sim}. Finally, the functions \texttt{plot.Musshoff.Hirschauer.group.means} and \texttt{plot.Karlan.List.group.means} are implemented by directly plotting the data provided by the supplementary material of our two applications, without involving any simulations or other functionality of the \texttt{asd} package.

The functions \texttt{Musshoff.Hirschauer.rejection.table} and \texttt{Karlan.List.rejection.table} may be used to produce the data presented in Tables \ref{tab:musshoff} and  \ref{tab:karlan_list}, respectively. These functions are implemented by calling \texttt{single.stage.rejection.table} and \texttt{two.stage.rejection.table}, which in turn calls \texttt{single.stage.game.test} and \texttt{two.stage.game.test} repeatedly for different total sample sizes.

The most important functions used from the \texttt{asd} package are \texttt{treatsel.sim}, \texttt{dunnett.test} and \texttt{hyp.test}. \texttt{treatsel.sim} is the main function called by \texttt{two.stage.sim} and performs most of the actual simulation work. The functions \texttt{dunnett.test} and \texttt{hyp.test} are used by \texttt{single.stage.game.test} and \texttt{two.stage.game.test} to perform the appropriate Dunnett tests for the intersection hypotheses and to combine the results into an overall result for deciding which individual hypotheses that could be rejected. For further information about these \texttt{asd} functions, see the vignette accompanying the latest version of the package, which can be found at \url{https://cran.r-project.org/web/packages/asd/asd.pdf}.

\end{document}